\documentclass[11pt, oneside]{article} 
\pdfoutput=1
\usepackage{geometry} 
\usepackage{graphicx}
	
\usepackage{amssymb}
\usepackage{amsmath}
\usepackage{tikz}
\usepackage{placeins}
\usepackage[final]{pdfpages}
\usepackage{commath}
\usepackage{braket}
\usepackage{subcaption}
\usepackage{float}

\usepackage{hyperref}
\usepackage{xcolor}
\hypersetup{
  colorlinks   = true, 
  urlcolor     = red, 
  linkcolor    = blue, 
  citecolor   = blue 
}

\usepackage{csquotes}
\usepackage[
    style=numeric-comp,
    bibstyle=phys,
    sorting=none,
    maxnames = 5,
    biblabel=brackets,
    backend=biber,
    eprint=true
]{biblatex}
\newcommand{\me}{\mathrm{e}}
\newcommand{\half}{\frac{1}{2}}
\DeclareRobustCommand{\orderof}{\ensuremath{\mathcal{O}}}
\DeclareRobustCommand{\exp}{\ensuremath{\mathrm{e}}}
\DeclareRobustCommand{\d}{\ensuremath{\mathrm{d}}}
\DeclareMathOperator\arctanh{arctanh}
\newcommand*\Eval[3]{\left.#1\right\rvert_{#2}^{#3}}

\title{\textbf{Exploring ALPs beyond the canonical}}
\author{Gonzalo Alonso\,-{\'A\!}lvarez and Joerg Jaeckel \vspace{3mm} \\  \textit{ \normalsize Institut f{\"u}r Theoretische Physik, Universit{\"a}t Heidelberg,}\\ \textit{\normalsize Philosophenweg 16, 69120 Heidelberg, Germany}}
\date{}

\begin{document}
\maketitle

\begin{abstract}
Axion-like particles (ALPs) are interesting dark matter candidates both from the theoretical as well as from the experimental perspective. 
Usually they are motivated as pseudo-Nambu-Goldstone bosons. 
In this case one of their most important features is that their coupling to other particles is suppressed by a large scale, the vacuum expectation value of the field breaking the symmetry that gives rise to them. 
This naturally endows them with very weak interactions but also restricts the maximal field value and therefore the regions where sufficient dark matter is produced. 
In this paper we investigate deviations from this simplest setup, where the potential and interactions are as expected for a pseudo-Nambu-Goldstone boson, but the kinetic term has singularities. 
This leads to a significantly increased area in parameter space where such particles can be dark matter and can be probed by current and near future experiments. 
We discuss cosmological limits and in the course of this give a simple derivation of a formula for isocurvature fluctuations in models with general anharmonic potentials.
As an application of this formula we give an update of the isocurvature constraints for QCD axion dark matter models, using the most recent results for the QCD topological susceptibility and the newest Planck data.
\end{abstract}

%\tableofcontents

\section{Introduction}\label{sec:Bounds}

Axions and axion-like particles (ALPs) are a prediction of some of the best-motivated beyond the standard model physics scenarios (see, e.g.~\autocite{jaeckel_low-energy_2010,cicoli_axion-like_2013,ringwald_axions_2014} for reviews).
Many of their properties are determined by two quantities: the mass, $m$ and the so-called decay constant, $f_a$.
An important feature that all these particles share is that they enjoy a shift symmetry, a discrete version of which is preserved at the quantum level.
The existence of this symmetry protects their potential from quantum corrections that could otherwise be very large.
In the framework of quantum field theory, such particles arise as pseudo Nambu-Goldstone bosons of approximate global chiral symmetries~\autocite{peccei_cp_1977,weinberg_new_1978,wilczek_problem_1978,masso_light_1995,masso_new_1997,masso_planck-scale_2004}.
In other setups such as supergravity or string theory, particles with similar properties appear in the spectrum.
For instance, ALPs are a general consequence of the compactification of extra dimensions and string theory~\autocite{svrcek_axions_2006,douglas_flux_2007,arvanitaki_string_2010,acharya_m_2010,higaki_note_2011,marsh_axiverse_2011,cicoli_type_2012}.
In that context, there can be dozens of such particles whose potentials, kinetic terms and interactions may contain a large number of free parameters.
In an attempt to accommodate all these similar particle candidates, we will talk about ALPs in the general sense of a light (pseudo-)scalar particle, and we will reserve the term ``axion" to refer to ALPs that couple to the gluon field strength tensor through the QCD topological term and solve the strong CP problem.

Axion-like particles are excellent candidates to account for some or all the dark matter that we observe in the universe~\autocite{arias_wispy_2012,ringwald_exploring_2012,jaeckel_family_2014,marsh_axion_2016}.
Cosmological and astrophysical observations tell us that dark matter particles should be weakly interacting, stable at cosmological scales and cold.
ALPs can naturally fulfil all these requirements.
First, the discrete shift symmetry constrains their possible couplings to other fields, and those that are allowed are typically suppressed by $f_a$, which can be a large energy scale.
This fact, together with their small mass which limits the possible number and type of decay products as well as the phase space, makes them extremely stable.
Naively, the fact that they are very light might seem to contradict the requirement that the ALP dark matter population should be cold.
However, it is easy to see that this is not necessarily the case.
Because of their feeble interactions with other particles, ALPs are not produced thermally, but rather by the so-called misalignment mechanism, which yields a very non-relativistic population of ALPs that behave as cold dark matter~\autocite{preskill_cosmology_1983,abbott_cosmological_1983,dine_not-so-harmless_1983,arias_wispy_2012,jaeckel_family_2014}.

All in all, ALPs and axions are well motivated dark matter candidates, but their possible mass and decay constant span many orders of magnitude thereby providing a significant challenge for experimental tests.
Fortunately, their properties, in particular their low mass, also provides for new opportunities for experimental searches and theoretical arguments that can be used to probe their parameter space (see~\autocite{graham_experimental_2015} for a recent review).

\begin{figure}[t!]
\centering
\includegraphics[width=0.625\textwidth,height=\textheight,keepaspectratio]{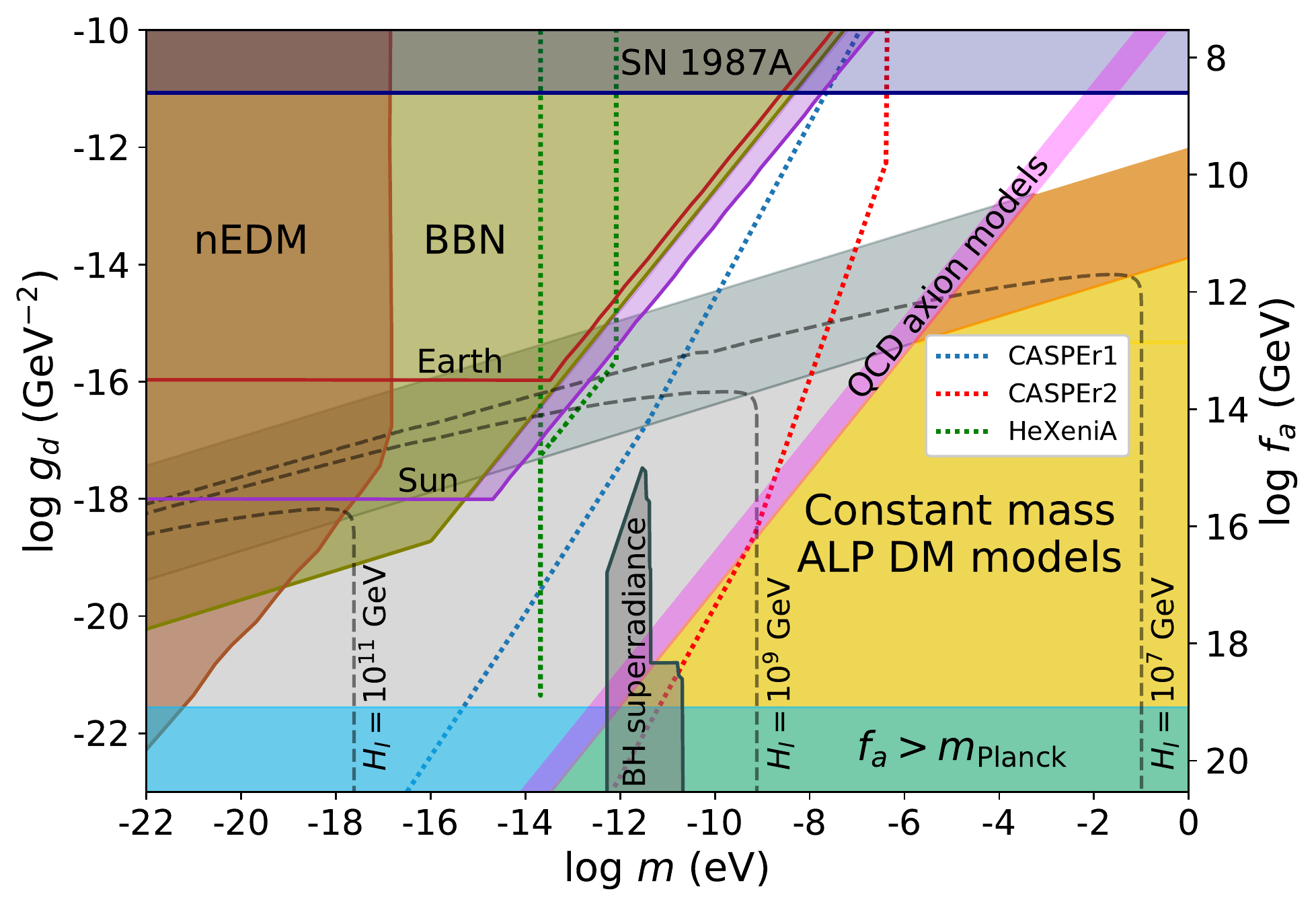}
\caption{Parameter space for canonical axion-like particles, considering gravitational effects and interactions derived from the QCD $G\tilde{G}$ term.
On the horizontal axis we plot the mass of the ALP, while the vertical axis gives the decay constant $f_a$ on the right and the effective coupling to nucleons $g_d\propto f_a^{-1}$ on the left.
Canonical ALP models with a constant mass can only generate enough dark matter via the misalignment mechanism in the yellow and grey shaded areas.
Accounting for the anharmonicities of the potential and allowing for a fine-tuned initial condition, this region can be enlarged to also include the orange band (we take the lowest viable Hubble scale of inflation, $H_I\sim 4.5\cdot 10^{-23}$ GeV).
Note that the QCD axion models are restricted to lie on the magenta line.
Taking the interaction to be given by Eq.~\eqref{eq:coupling}, the region to the left of the QCD axion line is disfavoured by the unavoidable (temperature dependent) contribution to the mass from QCD effects~\autocite{blum_constraining_2014} (see also \secref{sec:QCD}).
This region is shown in light grey.
The dark blue region is excluded by the supernova limits estimated in~\autocite{raffelt_astrophysical_2008}. 
Shaded in brown is the area where experiments looking for a static nuclear electric dipole moment (nEDM, see~\autocite{abel_search_2017}) would have found the oscillating one, while the dotted lines represent sensitivity estimates for future oscillating EDM experiments~\autocite{budker_cosmic_2014,hexenia_2017}.
In the dark green region ``BBN'' ALPs coupled to QCD are inconsistent with the production of the observed abundance of light elements during Big Bang Nucleosynthesis \autocite{blum_constraining_2014}.
The violet and dark red lines dubbed ``Earth" and ``Sun" correspond to constraints from the ALP field being sourced by dense astrophysical objects~\autocite{hook_probing_2017}.
The dark grey area is disfavoured by the observation of quickly rotating stellar black holes which would have been spun down in a superradiant process (from~\autocite{arvanitaki_discovering_2015}).
The area above the dashed black lines, plotted for different values of the Hubble scale of inflaton $H_I$, is disfavoured due to the generation of too much power in isocurvature perturbations at the scales probed by the Planck satellite~\autocite{ade_planck_2016} (see more details in~\secref{sec:Isocurvature_perturbations}).
Finally, $f_a$ is (softly) bounded from above by the requirement that it does not exceed the Planck scale.}
\label{fig:ALP_parameter_space}
\end{figure}

Experimental tests are usually dependent on the coupling to Standard Model particles.
One example is a coupling to two gluons,
\begin{equation} \label{eq:coupling}
{\mathcal{L}}\supset\frac{\alpha}{8\pi f_{a}}\phi G_{\mu\nu}\tilde{G}^{\mu\nu}.
\end{equation}
This coupling also induces a coupling to a nucleon electric dipole moment (EDM),
\begin{equation}
{\mathcal{L}}\subset g_{d}\phi\bar{N}\sigma_{\mu\nu}F^{\mu\nu}N
\end{equation}
that is particularly important for searches when $\phi$ is dark matter\footnote{The coupling~\eqref{eq:coupling} also induces tree-level $P,T$-violating forces between nucleons, which can give a larger contribution to atomic EDMs than the loop-induced nucleon EDMs~\autocite{stadnik_axion-induced_2014}. This is relevant for EDM experiments that use atoms instead of free neutrons, like some of the ones presented in~\figref{fig:ALP_parameter_space}. For those, the limits and projections should be understood as applying directly to $f_a$ and not $g_d$.}.
The coupling constants are related via~\autocite{pospelov_theta-induced_1999,graham_new_2013}
\begin{equation}
g_{d}\approx \frac{2.4\cdot 10^{-16}}{f_a}\ \mathrm{e}\cdot\mathrm{cm} \approx 3.4\cdot 10^{3} \mathrm{GeV}^{-2} \left(\frac{\mathrm{GeV}}{f_{a}}\right).
\end{equation}

\figref{fig:ALP_parameter_space} summarises the constraints that can be cast on the canonical ALP dark matter scenario from these interactions with the visible sector.
In addition we show limits that arise from unavoidable gravitational interactions.

Unfortunately, some of the theoretically favoured existing models require high decay constants for the ALPs to be able to account for all the dark matter energy density that we observe in our Universe.
This means that some of the better motivated combinations of $(m,f_a)$ are not in the best position to be tested, be it through gravitational interactions or through couplings to gluons and nucleons or photons.
It is therefore timely to search for models that can accommodate low enough values of $f_a$ that can be in reach of these searches, while still being able to produce the required dark matter abundance. One option is to enlarge the field range by a monodromy~\autocite{silverstein_monodromy_2008,mcallister_gravity_2010,kaloper_natural_2009} as done in~\autocite{jaeckel_monodromy_2016}. 

In this paper we pursue the same goal by employing a non-standard kinetic term for the ALP field.
This is a possibility that has been exploited in the literature~\autocite{alishahiha_dbi_2004,domcke_pbh_2017} in the context of inflationary models (though not so much for axion inflation), but to our knowledge such a study has not been performed for dark matter models.
As we will see, a very rich phenomenology arises when this possibility is allowed.
Of special interest is that this scenario will indeed be able to populate regions of the parameter space that can be tested in the near future, either with astrophysical observations or experimental searches.
Focusing on the coupling to nucleons, the main motivation for us in this respect is threefold. 
First, as was already argued, we want to explore the possibility of building an ALP dark matter model with a larger such coupling.
Second, we ask ourselves if these models could lie on the region of parameter space to the left of the QCD axion band in~\figref{fig:ALP_parameter_space}.
Finally and concerning the Big Bang Nucleosynthesis bound that seems to restrict this area of parameter space, we would like to test its robustness constraining such ALP models.

In this work we study the viability of ALPs with a non-canonical kinetic term as dark matter candidates from a purely phenomenological perspective. 
Let us nevertheless briefly mention some of the mechanisms that can give rise to this scenario.
For instance, a non-minimal coupling of the ALP field to gravity in the so-called Jordan frame induces a non-canonical kinetic function in the usual Einstein frame.
In the context of supergravity, an explicit breaking of the shift symmetry in the K{\"a}hler potential also results in non-standard kinetic terms for the ALP.
Finally, in the context of compactifications, string theory \textit{a priori} contains all the necessary ingredients to generate axions with non-canonical kinetic terms, caused, for example, by back-reaction effects.
However, no explicit construction of the models that we consider in this work exists as of today, and this task is beyond the scope of this paper.
We leave the study of the possibility of embedding this phenomenological study in a more complete framework for future work.

This paper is structured as follows: in~\secref{sec:NonstandardKT} we discuss the effects of non-canonical kinetic terms and set up our explicit case study. In~\secref{sec:Cosmological_evolution} we study how this modified kinetic terms affects the cosmological evolution of the ALP field, and in~\secref{sec:Isocurvature_perturbations} we analyse the isocurvature perturbations predicted in this setup. In~\secref{sec:QCD} we discuss the impact of allowing for a coupling to QCD in this scenario, and conclude in~\secref{sec:Conclusions}.

Before getting started on the details we note that, although in this paper we focus mainly on the example of gluon interactions, most of our discussion is completely general and can be applied to any other coupling. 
Moreover, while the structure of interactions that we consider is inspired by that of pseudo-Nambu-Goldstone bosons, the essential qualitative features should also apply in the case of more general scalars and only depends on the singularities of the non-canonical kinetic terms.

%%%%%%%%%%%%%%%%%%%%%%%%%%%%%%%%%%%%%%%%%%%%%%%%%

\section{Non-canonical kinetic terms}\label{sec:NonstandardKT}

In this section we examine the effect that a non-standard kinetic term can have on the dynamics of the ALP field. 
Let us start with the Lagrangian
\begin{equation}\label{eq:basic_lagrangian}
{\cal L} = \frac{1}{2} K^2(\phi)\partial^\mu\phi\partial_\mu\phi - V(\phi),
\end{equation}
where we have allowed for a general real scalar (and positive definite) function of $\phi$, $K^{2}(\phi)$,  to scale the kinetic term and thus render it not canonically normalised. 
For definiteness, we will work with the usual periodic potential for ALP fields,
\begin{equation}\label{eq:potential}
V(\phi) = \Lambda^4 \left( 1 - \cos \frac{\phi}{f_a} \right).
\end{equation}
We now proceed by performing a field redefinition to obtain the canonically normalised field. 
The formal solution is to define
\begin{equation}\label{eq:field_redefinition_general}
\varphi (\phi) = \int K(\phi) d\phi \equiv g(\phi),
\end{equation}
and thus the Lagrangian for $\varphi$ is
\begin{equation}
{\cal L}(\varphi) = \frac{1}{2} \partial^\mu\varphi\partial_\mu\varphi - V(g^{-1}(\varphi)).
\end{equation}
Being canonically normalised, $\varphi$ is the physical (propagating) field. 
Let us see what kind of functions $K$ result in $\varphi$ being a viable dark matter candidate. 

The first condition is that $\varphi$ behaves like cold dark matter in the late universe. This requires that it oscillates harmonically at late times (see, e.g.~\autocite{arias_wispy_2012}).
Accordingly the kinetic term should not modify the dynamics close to the origin. 
This is automatic if the kinetic term approaches a non-vanishing constant value close to the origin,
\begin{equation}
K\rightarrow {\rm const.}=1\quad{\rm for}\quad \varphi\rightarrow 0.
\end{equation}
As indicated in the equation, this constant can be chosen to be equal to $1$ by a suitable choice of normalisation.

So why should we now choose a non-trivial function for $K$ and what shall we choose?
As already mentioned in the introduction, we would like to find a model with larger couplings, i.e. smaller $f_{a}$, that still gives a sufficient dark matter density.
Roughly speaking the problem of obtaining a sufficient energy density can be understood as follows. For the potential Eq.~\eqref{eq:potential} the maximal initial energy density is given
by $\Lambda^4$. This is linked to the mass $m$ of the particle via $\Lambda^4=m^2f^2_{a}$.
If $f_{a}$ is too small the initial and in consequence the final energy density is too small to make up all of the dark matter.

One way to avoid this problem would be to break the periodicity of the potential~\eqref{eq:potential} such that the potential continues to grow for large field values, e.g. by exploiting a monodromy~\autocite{jaeckel_monodromy_2016}.

\begin{figure}[t!]
\centering
\includegraphics[width=\textwidth/2,height=\textheight,keepaspectratio]{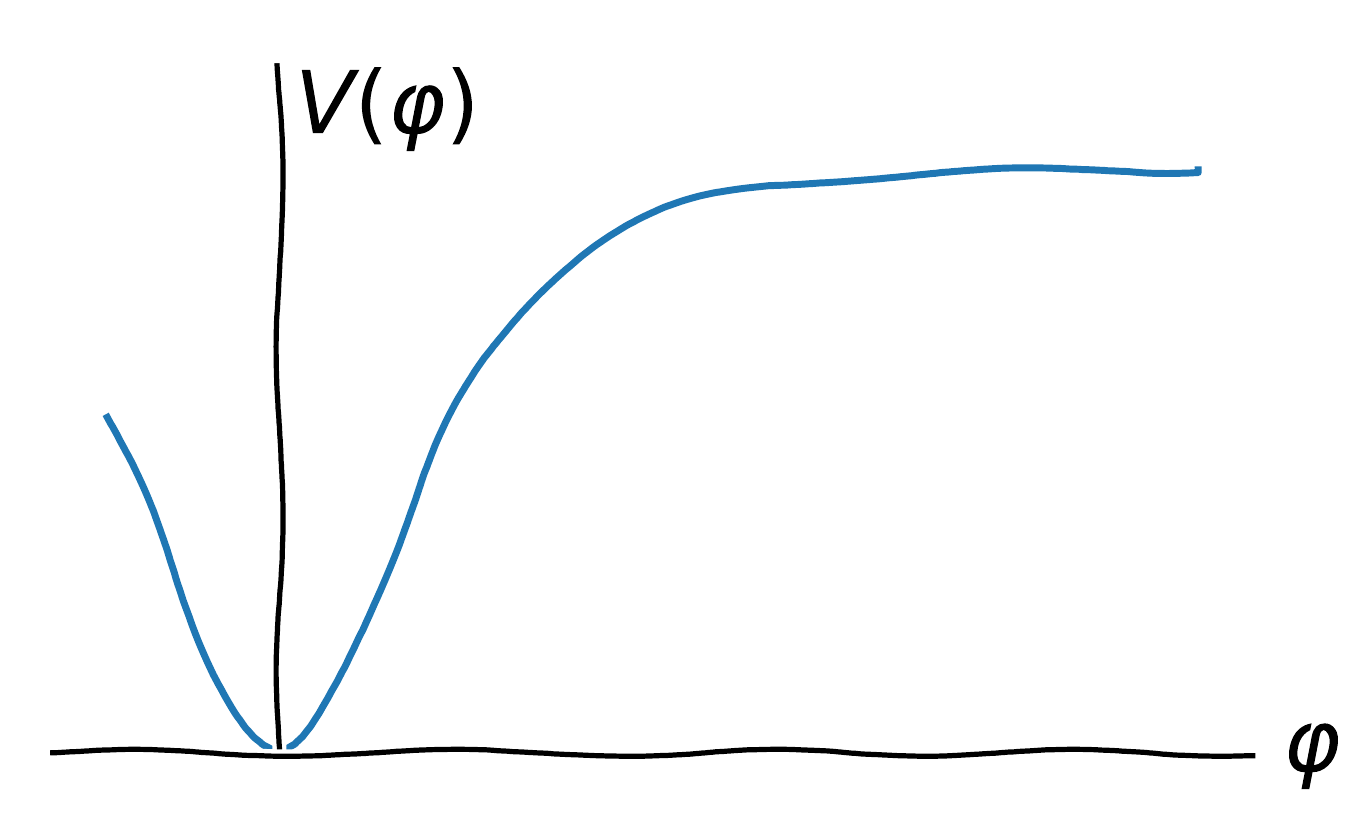}
\caption{Slow roll-like potential. }
\label{fig:SR_potential}
\end{figure}

Here we will explore a different  strategy. As long as the Hubble constant is sufficiently large the evolution of the field is frozen and the energy density is approximately constant. 
As discussed below the evolution and consequently the dilution of the energy starts when $H^2\sim |V''(\varphi)|$. Hence, we can increase the energy density today by choosing the kinetic function $K$ such that the potential becomes very flat for large field values\footnote{An alternative is to start in a region of field space where $V'(\varphi)$ is very small, i.e. the field is close to a maximum. However, this is strongly limited by the existence of inflationary fluctuations~\autocite{wantz_axion_2010,di_cortona_qcd_2016} (see also~\secref{sec:Isocurvature_perturbations}).}. A cartoon of this is shown in~\figref{fig:SR_potential}.

Using
\begin{equation}
\frac{\partial V}{\partial \varphi} = \frac{\partial V}{\partial \phi} \cdot \frac{\partial \phi}{\partial \varphi} = \frac{1}{K} \frac{\partial V}{\partial \phi},
\end{equation}
we see that this can be easily achieved if $K$ has a singularity at some field value $\phi=a$,
\begin{equation}
K\rightarrow\infty\quad{\rm for}\quad \varphi\rightarrow a.
\end{equation}

This singular structure has an additional advantage: 
The non-canonically normalised field $\phi$ will never exceed $\phi=a$ during its evolution. 
Limits such as the one discussed in~\secref{secbbn} arising from BBN that are based on a sizeable field value at some earlier epoch can thus be avoided if $a$ is sufficiently small.

A simple function that satisfies the above requirements while keeping the periodic properties intact is,
\begin{equation}\label{eq:non_canonical_kinetic_term}
K(\phi) = \frac{1}{\cos \left( \frac{N\phi}{f_a} \right)} .
\end{equation}

While this choice might seem rather arbitrary at first, there are some arguments that make it more general than it seems.
The approach for obtaining a flattened potential for a scalar via a non-canonical kinetic term has been widely used in the context of inflationary cosmology~\autocite{alishahiha_dbi_2004,domcke_pbh_2017}. 
Indeed over the last years, $\alpha$-attractor models~\autocite{kallosh_superconformal_2013,kallosh_planck_2015} have attracted special attention.
In this context, \autocite{galante_unity_2015} showed that the determining property of this class of models is the existence of a pole in the kinetic term.
More precisely, it is the order and the residue of the pole that play a key role, and not so much the precise functional form of the kinetic function.
We can therefore be confident that our results will not depend much on the specific choice of $K$.
Similarly to~\autocite{galante_unity_2015}, here we focus on the case of a second-order pole.
As we mentioned before, this case is better motivated and may arise, for instance, as a consequence of a non-minimal coupling to gravity.
Nevertheless, we check in Appendix~\secref{sec:Appendix1} that our main conclusions remain unchanged if we allow for higher-order poles.

Also, recall that the shift symmetry $\phi\rightarrow \phi + \text{\textit{const.}}$ of the ALP field is what protects its mass from large corrections.
It thus seems sensible to preserve or only slightly break this symmetry.
Indeed, by our choice of potential Eq.~\eqref{eq:potential}, we are assuming that a small explicit breaking is present.
This breaking typically occurs at the nonperturbative level~\autocite{peccei_strong_2008, kim_axions_2010,} and crucially preserves the discrete shift symmetry $\phi\rightarrow \phi + 2 k \pi f_a$, which allows us to retain a sufficient level of protection against quantum corrections.
We would like the kinetic term to preserve, at least, this discrete shift symmetry, which requires that $K(\phi)$ is a periodic function of $\phi/f_a$.
These arguments quickly lead us to Eq.~\eqref{eq:non_canonical_kinetic_term}.
Once again, we stress that the fact that we are writing a specific kinetic term should not be understood as a construction of a complete model, bur rather as a benchmark for our phenomenological study.

\bigskip
The transformation to the canonically normalised field is given by
\begin{equation}\label{eq:field_redefinition}
\varphi(\phi) = \frac{2f_a}{N} \arctanh \left( \tan \frac{N\phi}{2f_a} \right). 
\end{equation}
We should note that the poles of $K(\phi)$ are located at $\phi/f_a=\pi/(2N)$.
This means that, when doing the field redefinition \eqref{eq:field_redefinition}, we are restricting the field space to $\phi/f_a\in (- \frac{\pi}{2N}, \frac{\pi}{2N})$.
As already mentioned above this will become important when discussing the limits arising from a gluon coupling in~\secref{sec:QCD}.
In principle there exist a total of $N$ different branches $\phi/f_a\in \left( (k-\half)\frac{\pi}{N}, (k+\half)\frac{\pi}{2N}\right)$ where the field could be trapped.
However, the only one which has a minimum in the potential is the one closest to the origin.
In other branches, the field would slow-roll towards infinity\footnote{In principle one could have tunnelling between different branches. If the decay time of the metastable vacuum is small enough, the field would always eventually end up in the branch closest to zero. However, a calculation of the tunnelling rate is highly model dependent and beyond the scope of this work.}, making them unappealing for the phenomenologically purposes that the we have in mind.
For this reason, we focus on the phenomenologically viable region around zero.

\begin{figure}[t!]
\centering
\includegraphics[width=0.7\textwidth,height=\textheight,keepaspectratio]{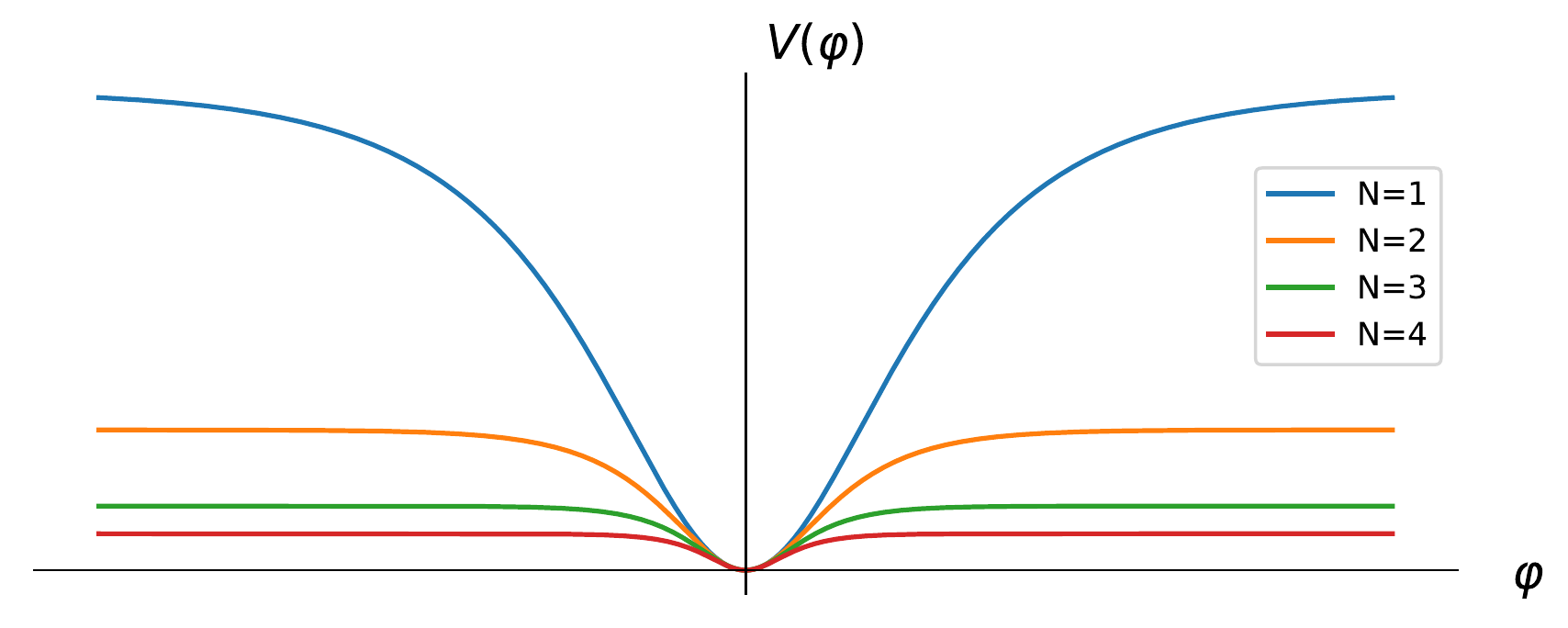}
\caption{Potential for the canonically normalised field, plotted for various values of $N$.
 Note that the potential is quadratic for small field value but flattens away from the origin.}
\label{fig:potential}
\end{figure}

Using the field redefinition~\eqref{eq:field_redefinition} the Lagrangian for the canonically normalised field is given by
\begin{equation}\label{eq:canonically_normalised_Lagrangian}
{\cal L} =   \frac{1}{2} \partial^\mu\varphi\partial_\mu\varphi - \Lambda^4 \left[ 1 - \cos \left( \frac{2}{N} \arctan \left( \tanh \frac{N\varphi}{2f_a} \right) \right) \right].
\end{equation}
By expanding about the origin, it can be checked that we indeed recover the quadratic behaviour for small field values.
The potential is plotted in~\figref{fig:potential} for different values of $N$.
It indeed looks quite similar to what we imagined in~\figref{fig:SR_potential}.

What about the equations of motion?
Let us assume that we have a homogeneous and isotropic field, $\phi = \phi (t)$ and consequently $\varphi = \varphi (t)$. 
The Klein-Gordon equation for a homogeneous and isotropic field in an expanding spacetime is
\begin{equation}
\ddot{\varphi} + 3H\dot{\varphi}+\partial_\varphi V(\varphi) = 0,
\end{equation}
where $H$ is the Hubble expansion parameter.

For convenience we introduce the dimensionless field variable,
\begin{equation}
\psi = \varphi / f_a, 
\end{equation}
in analogy to how the $\theta$ angle relates to the original axion field. 
Thus, we will be expressing the field value in terms of $f_a$ units. 
The equation of motion can then be written as
\begin{equation}
\ddot{\psi} + 3H \dot{\psi} + m^2 \frac{1}{\cosh N\psi} \sin \left[ \frac{2}{N} \arctan \left( \tanh \frac{N\psi}{2} \right) \right]=0,
\end{equation}
where we define 
\begin{equation}
m^2 = \frac{\Lambda^4}{f_a^2},
\end{equation}
which corresponds to the second derivative of the physical field around the minimum at 
$\psi=\varphi=\phi=0$. $m$ is the physical mass of the dark matter particles. 

%%%%%%%%%%%%%%%%%%%%%%%%%%%%%%%%%%%%%%%%%
\newpage
\section{Cosmological evolution and dark matter production}\label{sec:Cosmological_evolution}

The goal of this section is to find an estimate for the dark matter density in the model defined above and compare it with the observed abundance. 
The energy density of the field depends on the parameters $(f_a, m, N)$, as well as the initial conditions for the field and its cosmological evolution. 
For this purpose it is useful to briefly recall the misalignment mechanism~\autocite{preskill_cosmology_1983,abbott_cosmological_1983,dine_not-so-harmless_1983}, which gives us the basic idea of how our field evolves in a cosmological setup.

\subsection{The misalignment mechanism}\label{sec:misalignment}

Here we briefly summarise how a misaligned light scalar field evolves in an expanding spacetime, closely following the description in~\autocite{arias_wispy_2012}. 
Let us consider the simplified case of a real scalar field with Lagrangian
\begin{equation}
{\cal L} = \frac{1}{2}\partial_\mu \phi \partial^\mu \phi - \frac{1}{2}m_\phi^2\phi^2.
\end{equation}
Note that our final goal is not the harmonic case but a more complicated potential with strong anharmonicities.
However, solving this simplified equation will give us helpful insights to tackle the anharmonic potential.
In a homogeneous setting, the equation of motion for $\phi$ is
\begin{equation}\label{eq:simpleeom}
\ddot{\phi} + 3H\dot{\phi} + m_\phi^2 \phi = 0.
\end{equation}
This is the equation of a damped harmonic oscillator. 
There are two distinct regimes in the evolution of $\phi$. 
First, at very early times when $3H \gg m_\phi$, the oscillator is overdamped and so the solution is $\dot{\phi} = 0$, and the field is stuck at its initial value. 
At a later time $t_1$ such that $3H (t_1) = m_\phi$, the damping has decreased enough so that the field can start to oscillate.
The equation of motion for the oscillating regime can then be solved using the WKB approximation:
\begin{equation}\label{eq:harmonic_WKB}
\phi(t) \simeq \phi(t_1) \left( \frac{a(t_1)}{a(t)} \right) ^{3/2} \cos \left( m_\phi (t-t_1) \right),
\end{equation}
where $a(t)$ is the scale factor.
We see that the energy density, which is proportional to the amplitude of the oscillations squared, dilutes with expansion as $a^{-3}$. 
This means that the oscillating field behaves like pressureless matter for all processes mediated by gravitation.
In this simplified setup, the energy density in the axion field today is
\begin{equation}\label{eq:harm_energy_density}
\rho_\phi (t_0) \simeq 0.17 \frac{\text{keV}}{\text{cm}^3}\ \sqrt{\frac{m_\phi}{\text{eV}}} \left( \frac{\phi_0}{10^{11}\ \text{GeV}} \right)^2 {\cal F}(T_1),
\end{equation}
where 
\begin{equation}
{\cal F}(T_1)= \frac{\left( g_\star(T_1) / 3.36 \right)^{\frac{3}{4}}}{\left( g_{\star S}(T_1) / 3.91\right)}
\end{equation}
is a smooth function (cf.~\autocite{arias_wispy_2012}) that varies from $1$ to $\sim 0.3$ when $T_1\in (T_0,200\ \text{GeV})$. 
The last result assumes that the field starts oscillating during radiation domination and that the comoving entropy is conserved.

\subsection{Analytical estimate of the dark matter density}\label{sec:analytic_estimate}

After this small detour to explain the misalignment mechanism for the harmonic potential, let us go back to our case of interest: the ALP field with a non-standard kinetic term. 
Recall that the equation of motion that we have obtained for the physical field $\psi$ is
\begin{equation}\label{eq:nonlinear_eom}
\ddot{\psi} + 3H \dot{\psi} + m^2 \frac{1}{\cosh N\psi} \sin \left[ \frac{2}{N} \arctan \left( \tanh \frac{N\psi}{2} \right) \right] = 0.
\end{equation}
We see that in the limit of small $\psi$, when $N\psi \ll 1$, this reduces to the simplified case \eqref{eq:simpleeom} and the evolution is exactly as we described in the simple real scalar field case. 
However, the situation is different in the regime $N\psi \gtrsim 1$. 
As we can expect by looking at~\figref{fig:potential}, the flatness of the potential away from the minimum at $\psi=0$ will have the effect of delaying the start of the oscillations. 
Moreover, the oscillations, once they start, will not be harmonic until the damping has made the amplitude decrease enough to be in the small field regime. 
This means that the WKB approximation might not be as good in this case.

Although we suspect that the WKB approximation might break down when the amplitude of the oscillations is big due to the anharmoniticity of the potential, we will use it as a first approximation to solve the equation of motion and get an analytical estimate of the result. We will later contrast this to a more precise numerical computation. 
In the analytical approach, we will study the two regimes, where the damping is over- and under-critical, respectively, and build up the global evolution of the field by glueing together the solution for each regime. 
Our goal is to compute the current energy density of dark matter-like particles given an initial condition for the physical field. 

As we saw, the first thing to do is to find the time when the oscillations start. 
In analogy with the simple case, where the condition was $3H = m_\phi$, we use a generalisation of this formula for a non harmonic potential, namely
\begin{equation}
3H = \left| V^{\prime\prime} (\psi_0) \right| ^{1/2}.
\end{equation}
In~\secref{sec:numerics} we will see that this indeed works reasonably well to determine when the oscillations start, as it takes into account the flatness of the potential away from the origin. 
In the limit of large $N\psi\gg 1$, the second derivative of the potential can be written as
\begin{equation}
V^{\prime\prime}(\psi) \simeq -2Nm^2 \exp^{-N\psi}\sin \frac{\pi}{2N}.
\end{equation}
This turns out to be a very good approximation for intermediate and even small values of $N\psi$.
One key difference with the harmonic case is that here the point in time when oscillations begin depends on the initial field value $\varphi_0$.
With this we already see that the oscillations are exponentially delayed for big $N\psi$:
\begin{equation}\label{eq:tstart}
t_s \equiv t_{\text{start}} = \frac{3}{2\left| V^{\prime\prime}(\psi_0) \right| ^{1/2}} \simeq \frac{3}{2m} \left( 2N\sin \frac{\pi}{2N} \right)^{-1/2} \exp^{\frac{N\psi_0}{2}} \propto \exp^{\frac{N\psi_0}{2}},
\end{equation}
where we have assumed radiation domination so that $H=1/(2t)$. 
We now use this as an initial condition for the WKB approximation. 
In this approximation, the energy density of the physical field $\varphi$ is
\begin{equation}\label{eq:energy_density_semianalytic}
\rho_\varphi (T) = \frac{1}{2} m^2 f_a^2 \psi_0^2 \frac{g_{\star S}(T)}{g_{\star S}(T_s)} \left( \frac{T}{T_S}\right)^3,
\end{equation}
where we have used the conservation of comoving entropy $S=sa^3$ to express it in terms of temperatures instead of scale factors. 
Using the expression for the Hubble constant during radiation domination 
\begin{equation} \label{eq:Hubble_temperature}
H(T) = 1.66\sqrt{g_\star(T)} \frac{T^2}{m_{\text{pl}}},
\end{equation}
we can express the current energy density of the field as a function of the initial condition $\psi_0$,
\begin{equation}
\rho_\varphi \simeq 0.17\ \frac{\text{keV}}{\text{cm}^3} \cdot \sqrt{\frac{m}{1\ \text{eV}}} \left( \frac{f_a}{10^{11}\ \text{GeV}} \right)^{2} \psi_0^2\ {\cal F}(T_s) \cdot \left( 2N \sin \frac{\pi}{2N} \right)^{-3/4} \exp^{\frac{3}{4}N\psi_0}.
\end{equation}
We can compare this density with the one corresponding to a harmonic potential. 
The result is
\begin{equation}\label{eq:enhancement_analytic}
\frac{\rho^{\text{anh}}}{\rho^{\text{harm}}} \simeq \frac{\mathcal{F}(T_1)}{\mathcal{F}(T_s)} \cdot \left( 2N \sin \frac{\pi}{2N} \right)^{-3/4} \exp^{\frac{3}{4}N\psi_0} \sim \exp^{\frac{3}{4}N\psi_0},
\end{equation}
so the energy density is exponentially enhanced\footnote{In Appendix~\secref{sec:Appendix1} we check that a significant enhancement also exists if we allow for a kinetic function with a higher-order pole.} for large $N$ and initial condition $\psi_0$. 
The precise exponent that we obtain here should be taken as a very rough estimate.
Indeed, a numerical computation is needed to get a precise result, which is what we will aim for in the following section.

As we can see in \eqref{eq:enhancement_analytic} the enhancement is exponential in $N\psi$. This implies that the field values required to yield the correct dark matter abundance are usually not too large. In the phenomenologically interesting region we usually do not need to have values for $N\psi$ that are bigger than $50$. 
The largest initial field values happen for $N=1$ and are of order $50$ in units of $f_a$.

Another constraint that we have to care about is that the field is behaving like dark matter once it comes to dominate the dynamics of the universe, i.e. we do want to avoid having an additional phase of inflationary expansion driven by $\psi$.
A sufficient condition for this is that the field has already started to oscillate at matter radiation equality. 
Making use of the more precise numerical estimate that we will obtain in the next section, we can estimate what region of parameter space satisfies this condition,
\begin{equation}\label{eq:second_inflation_limit}
f_a \gtrsim 10^{-6}\ \mathrm{GeV}\cdot N\cdot \left( \frac{\mathrm{eV}}{m} \right)^{0.81}.
\end{equation}
This condition excludes the very small values of the mass and the decay constant in the upper left corner of \figref{fig:ALP_parameter_space}, which are already in tension with the nEDM experiment, BBN observations and the limits from \autocite{hook_probing_2017}.

\subsection{Numerical computation}\label{sec:numerics}

Having obtained a simple estimate of the cosmological evolution of the field, we now make use of a numerical solution of the equation of motion to have a more precise result.
Our goal in this subsection is to quantify how much the solution for the nonlinear equation of motion \eqref{eq:nonlinear_eom} deviates from the harmonic case \eqref{eq:simpleeom}.

Following the usual practice for dealing with anharmonicities in the ALP potential (see~\autocite{turner_cosmic_1986, lyth_axions_1992, strobl_anharmonic_1994, bae_update_2008},~\autocite{diez-tejedor_cosmological_2017} has a slightly different definition), we use an effective parametrisation in terms of an anharmonicity function $f(\psi_0)$, such that
\begin{equation}
\rho^{\text{anh}} = f(\psi_0) \rho^{\text{harm}},
\end{equation}
where $\rho$ is the energy density of the ALP field, computed late enough when it is already behaving as cold dark matter.
This function only depends on the initial misalignment angle, and it should account for all the deviations from the harmonic solution.
This approach is normally used to account for departures from the quadratic potential in the usual axion and ALP models.
Our case is slightly different, mostly because we are dealing with an unbounded field range.
As a consequence, the usual functional form for $f(\psi_0)$ does not work here.
Guided by the result obtained in the analytical approximation, we work with the following ansatz for the anharmonicity function:
\begin{equation}\label{eq:anh_func_ansatz}
f(\psi_0) = \me^{b N \psi_0},
\end{equation}
where $b$ is a real parameter to be determined.
This ansatz accounts for the exponential enhancement in energy density that we have found analytically.
The normalisation needed is that $f(\psi_0) \rightarrow 1$ when $\psi_0 \rightarrow 0$, so as to recover the harmonic case in the small field limit.

\begin{figure}[t!]
\centering
\includegraphics[width=0.75\textwidth,height=\textheight,keepaspectratio]{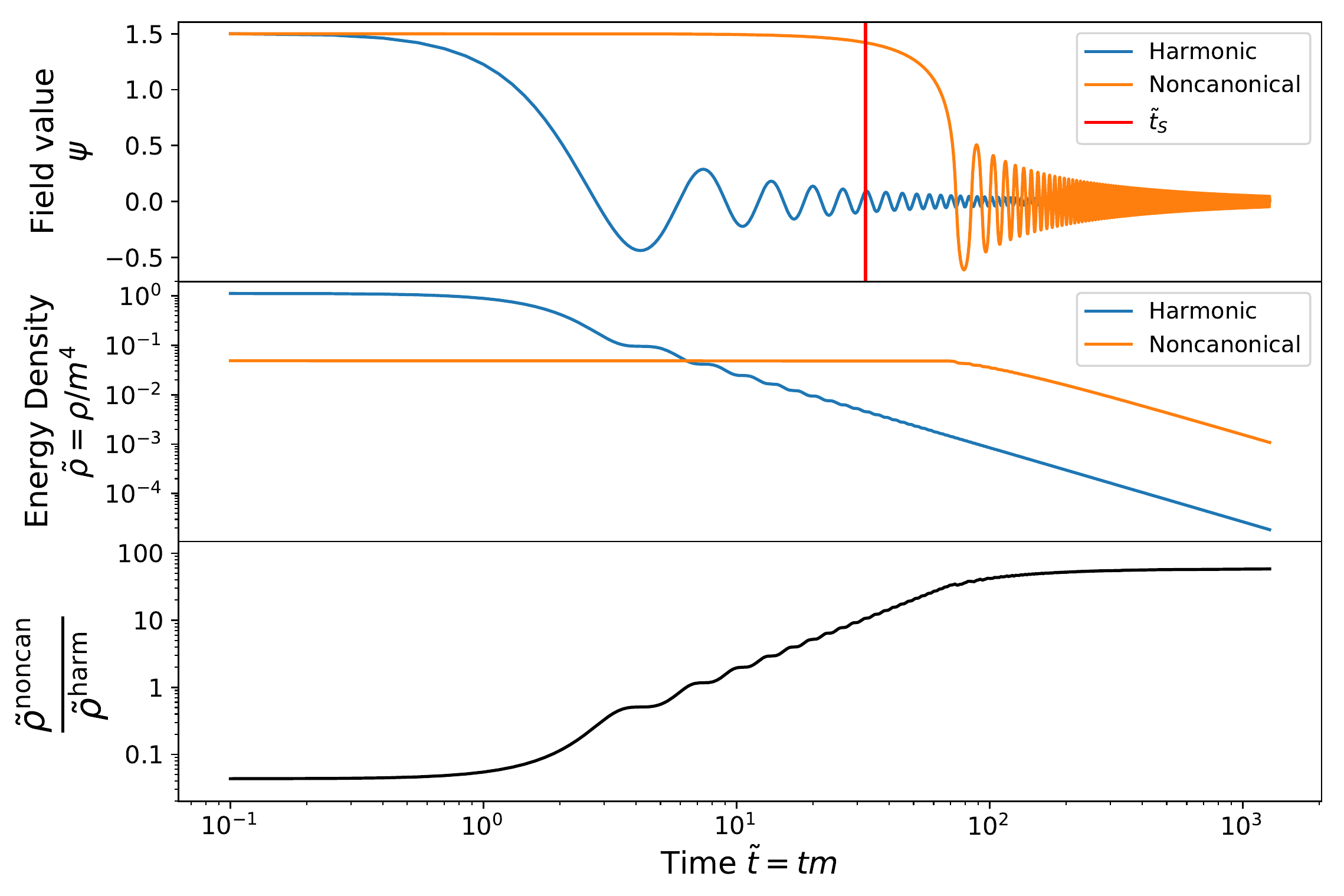}
\caption{Numerical solution of the non-canonical equation of motion compared to the harmonic solution, using $N=5$ and $\psi_0=1.5$ as an example. The top panel shows the solution for the field as a function of time, while the middle and bottom ones show the energy density of the field and the quotient of energy densities for the harmonic and non-canonical equations of motion. Note that this quotient approaches a constant as the adiabatic regime is reached, allowing us to obtain the anharmonicity factor. As a comparison and confirmation of our analytical results, the top panel also shows the time at which the oscillations are predicted to start in our analytical approach, Eq. \eqref{eq:tstart}.}
\label{fig:anh_function_evolution}
\end{figure}

The goal now is to fit the ansatz to a numerical computation of the energy density.
To set the problem in a more straightforward way, we want to compare the numerical solution of
\begin{equation}
\ddot{\psi} + 3\tilde{H}(\tilde{t}) \dot{\psi} + \tilde{m}^2 \frac{1}{\cosh N\psi} \sin \left[ \frac{2}{N} \arctan \left( \tanh \frac{N\psi}{2} \right) \right] = 0
\end{equation}
with the solution for the damped harmonic oscillator equation
\begin{equation}
\ddot{\psi} + 3\tilde{H}(\tilde{t}) \dot{\psi} + \tilde{m}^2 \psi = 0.
\end{equation}
In this computation we use dimensionless quantities measured in units of $m$, denoted with a tilde: $\tilde{H}, \tilde{t}, \tilde{m} \dots$
In these units, the time for the start of the oscillations in the harmonic case is $\tilde{t}_1^{\text{harm}} = 3/2$ (assuming radiation domination), and the period of the oscillations is $2\pi$.
We solve the equations numerically until we are well within the adiabatic regime in both cases (that is, when the amplitude of the oscillations has decreased enough so that the non-canonical potential is well approximated by the harmonic one).
Then, we compute the energy density $\rho = (1/2)f_a^2\dot{\psi}^2 + V(f_a\psi)$ and extract the anharmonicity factor as the quotient of both energy densities.
As we are within the adiabatic regime, $\rho$ scales as $\rho\propto a^{-3}$ in both cases, so the quotient will stay constant.
An example of the numerical solution can be seen in~\figref{fig:anh_function_evolution}.

\begin{figure}[t!]
\centering
\includegraphics[width=0.75\textwidth,height=\textheight,keepaspectratio]{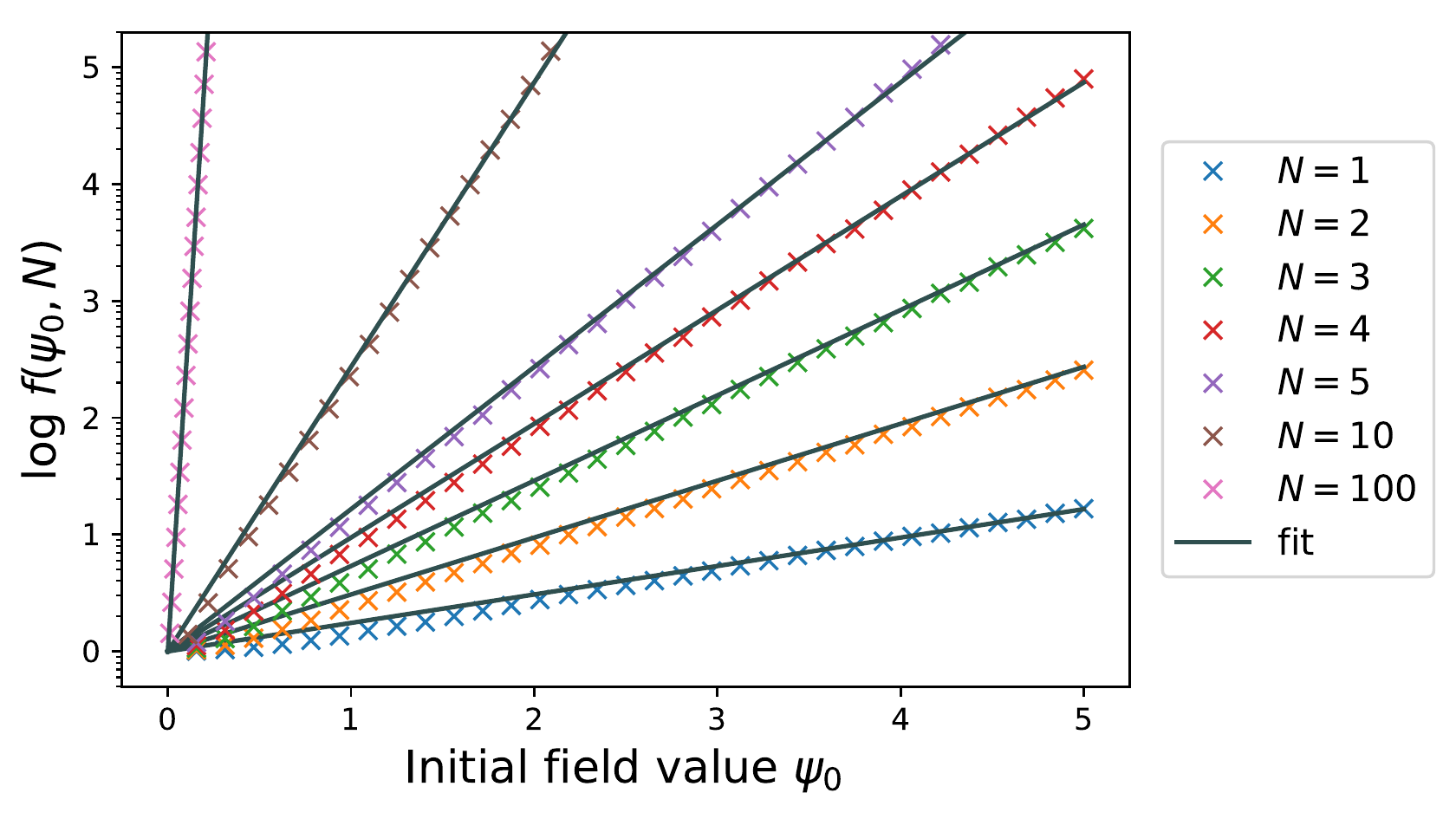}
\caption{Fit of the anharmonicity function to the ansatz in Eq. \eqref{eq:anh_func_ansatz}. We plot the result of the fit for a set of values of $N$ and a range of the initial misalignment angle $\psi_0\in(0,5)$.}
\label{fig:anh_function_fit}
\end{figure}

This process is repeated for a large number of values of $\psi_0$ and $N$ and we fit the results to the ansatz \eqref{eq:anh_func_ansatz}.
We obtain a very good fit with a value of $b = 0.56$, as can be seen in~\figref{fig:anh_function_fit}.
One should note that we are fitting a two dimensional data sample with just one parameter, so finding a good fit confirms that we have chosen an adequate ansatz.

The anharmonicity function allows us to compute the energy density of the non-canonical ALP field in a very simple way, combining the harmonic solution \eqref{eq:harm_energy_density} with the anharmonicity function~\eqref{eq:anh_func_ansatz}.
As long as we are within the adiabatic regime, the energy density in this approximation is given by
\begin{equation}\label{eq:anharm_energy_density}
\begin{aligned}
\rho_\psi^{\text{anh}} (t) &\simeq \frac{1}{2} f_a^2 m^2 f(\psi_0,N)\psi_0^2 \left( \frac{a_1^{\text{harm}}}{a(t)} \right)^3 \\
&= \frac{1}{2} f_a^2 m^2 f(\psi_0,N) \psi_0^2 \frac{g_{\star S}(T)}{g_{\star S}(T_1^{\text{harm}})} \left( \frac{T}{T_1^{\text{harm}}}\right)^3.
\end{aligned}
\end{equation}
The key difference between this equation and \eqref{eq:energy_density_semianalytic} is that here we use the well known solution of the harmonic equation of motion, instead of the full noninear one that arises in our non-canonical setup.
All the information about the nonlinearity is encoded in the anharmonicity function, making it much more manageable.

In the analytical approach, we found that the quotient between non-canonical and canonical density scales as $\rho_\text{NC}/\rho_\text{C}\sim \me^{(3/4)N\psi_0}$.
In the full numerical approach\footnote{In this study we have limited ourselves to the homogeneous field evolution.
Recently, the authors of~\autocite{soda_cosmological_2017} showed that potentials like the one we are considering can lead to a parametric resonance instability that can make inhomogeneous modes grow.
This effect may help to alleviate some tension that has been pointed out in~\autocite{irsic_first_2017} between the existence of ultralight ALPs and Lyman $\alpha$ forest observations.} we find a somewhat lower coefficient for the exponent of $0.56$.

We have seen that a non-canonical kinetic term can indeed enhance the energy density of ALP dark matter.
In the next few sections we will make use of the solutions for the cosmological evolution of the non-canonical ALP field to make predictions about its phenomenology, and to apply it to some particularly interesting cases.

%%%%%%%%%%%%%%%%%%%%%%%%%%%%%%%%%%%%%%%%%%%%%%%%%%%%%%%%%
\section{Isocurvature perturbations}\label{sec:Isocurvature_perturbations}

So far, we have assumed the initial misalignment angle $\theta_0 = f_a \phi_0$ to be a constant value all throughout the universe, but of course we have to take into account fluctuations, e.g. those imprinted by inflation. We do this by taking the initial misalignment angle as a spatially varying quantity, and describing it in terms of its average and variance.
Two very distinct scenarios arise, depending on whether the mechanism that gives rise to the ALP field turns on before or after the inflationary epoch of our Universe.

If the ALP field was established, e.g. by spontaneous symmetry breaking, after inflation, the variance of the angle can be large even within our Hubble volume.
The mean value will be $\phi_0 = 0$ and the energy density is given by the fluctuations as well as other effects such as, e.g. the decay of topological defects~\autocite{sikivie_cosmic_1991,hagmann_axion_2001}. In particular the latter contributions are not well understood and may also have some model dependence when going beyond the QCD axion.

To avoid this, we will focus on the scenario where the ALP field was present during inflation.
Classically, if the ALP field was established before inflation, then the spatial variance of the field within a Hubble patch will be washed out as spacetime is stretched during inflation.
This means that $\sigma_\phi^2 \rightarrow 0$, and the misalignment field can take any value $\phi_0$ in our Hubble patch.

However, this is not completely true, as any light field present during inflation will acquire quantum fluctuations (see, e.g.~\autocite{linde_particle_2005}).
The power spectrum of such fluctuations for a canonically normalised scalar field is scale invariant,
\begin{equation}
\braket{\left| \delta \phi (k) \right|^2} = \left( \frac{H_I}{2\pi} \right)^2 \frac{1}{k^3 / (2\pi^2)}.
\end{equation}
These fluctuations can be thought of as arising from a thermal spectrum at the Gibbons-Hawking temperature $T_{GH}=H_I/(2\pi)$~\autocite{gibbons_cosmological_1977}.
As long as these fluctuations do not restore the spontaneously broken symmetry that gives rise to the ALPs, i.e., as long as\footnote{It is also necessary that the symmetry is not restored during reheating~\autocite{beltran_isocurvature_2007}. We will assume this to be true.} $T_{GH}<f_a$, this will imprint small fluctuations on top of the otherwise homogeneous ALP field.
The corresponding fluctuations of the misalignment angle in Fourier space will have an amplitude of $\sigma_\phi (k) = H_I/(2\pi f_a)$.
In real space, the fluctuations are of a size $\sigma_\phi = \gamma H_I/(2\pi f_a)$, where $\gamma \sim \orderof (1)$ is a dimensionless factor that effectively encodes the dispersive effect of the logarithmically divergent small $k$ modes (see~\autocite{lyth_axions_1992}).
Its value depends on the length scales that we are interested in. Following~\autocite{hertzberg_axion_2008} we will set $\gamma = 2$ for the CMB characteristic scale $k_\star = 0.05\ \text{Mpc}^{-1}$.

As the ALP has a negligible contribution to the total energy density of the universe during inflation, fluctuations in the field do not contribute to the usual curvature perturbations.
Rather, they manifest themselves as fluctuations in the ratio of the number density of ALPs to the total entropy density, and are completely uncorrelated with the curvature perturbations.
This is the reason why they are called \textit{entropy} or \textit{isocurvature perturbations}.
As their interactions with other standard model particles are greatly suppressed, ALPs do not thermalise with the other species and their perturbations remain isocurvature~\autocite{weinberg_must_2004}.
At later stages of the cosmological evolution, the dark matter ALPs pick up a significant contribution to the energy density of the universe, and so they contribute to the temperature and polarisation fluctuations of the CMB as cold dark matter isocurvature modes.

Planck has set strong bounds on isocurvature perturbations~\autocite{ade_planck_2016},
\begin{equation}
\beta_\text{iso} = \frac{\Delta_\phi^2(k_\star)}{\Delta_\phi^2(k_\star) + \Delta_{\cal R}^2 (k_\star)} < 0.038
\end{equation}
at $95 \%$ CL.
Here, $\Delta_\phi^2(k_\star)$ and $\Delta_{\cal R}^2 (k_\star)$ are the power spectrum of the axion and curvature perturbations at the pivot scale $k_\star$, respectively.
Once the value of $\Delta_{\cal R}^2 (k_\star)$ is set (Planck gives $ \Delta_{\cal R}^2 (k_\star) = 2.1(9)\times 10^{-9} $), this translates into a bound on the axion isocurvature fluctuations.

To use this limits to constrain our scenario, we have to compute our prediction for
\begin{equation}\label{eq:Isocurvature_spectrum}
\Delta_\phi^2 = \Eval{\Braket{ \left( \frac{\delta\rho_\phi}{\rho_\phi}\right) ^2 }}{t_\text{CMB}}{},
\end{equation}
that is, we need to evolve the fluctuations in the energy density until the time of emission of the CMB and compare them with the homogeneous average value.

If the evolution of the field is linear, as it is in the case of canonical ALP models with a purely quadratic potential, the power spectrum is constant during the cosmological evolution.
As a consequence, one can evaluate it at any point, such as right after inflation and before the onset of the oscillations in the ALP field.
However, in any model that contains anharmonicities, the evolution at early times will be nonlinear, which implies that $\Delta_\phi^2$ will evolve nontrivially after inflation.
Thus, to arrive at the correct prediction for the isocurvature perturbations, we have to track the evolution of the fluctuations until late times.

In addition to the limits from isocurvature fluctuations,
the inflationary fluctuations\footnote{Quantum fluctuations of the ALP field should also be considered, but their effect is negligible when compared to the inflationary fluctuations.} in the ALP field also forbid tuning the initial misalignment angle with arbitrary precision.
In fact, there is an unavoidable limit to this tuning, and it is that our tuning precision cannot be better than the fluctuations, with $\sigma_\theta = \gamma H_I/(2\pi f_a)$, as was argued in~\autocite{wantz_axion_2010}.
This has two related consequences.
The first is that the initial misalignment angle cannot be infinitely close to zero.
The requirement that the current ALP energy density is not bigger than the measured dark matter density $\Omega_C h^2 \sim 0.12 $ then sets a bound on the parameter space.
This bound is model independent (as long as all the potentials are approximately quadratic for small $\theta$) and roughly requires
\begin{equation}
m < \left( \frac{10^{12} \text{ GeV}}{H_I} \right)^4 \text{ eV}.
\end{equation}
Secondly, if the field range is compact (as for the usual canonical ALP), an argument similar to the one above tells us that some regions of the parameter space will not yield enough energy density to account for all the dark matter.
Indeed, it is not possible to tune the initial value of the field at the top of the potential with infinite precision, due to the presence of fluctuations.
The requirement here is that $\pi - \theta_0 < \gamma H_I/(2\pi f_a)$.
This particular limit will strongly depend on the anharmonicity of the potential, so it is not possible to give a more explicit expression.
We discuss some particular cases in the next subsection.
However, this last effect will not be relevant in our non-canonical model, as there we have an unbounded field range (our potential does not have a maximum).

\subsection{Isocurvature perturbations for anharmonic potentials}\label{sec:Isocurvature_perturbations_anharmonic}

We now present a general analytical expression to compute the isocurvature perturbations in general ALP models where the potential might have big anharmonicities.
We do this using the anharmonicity function formalism that we presented in the previous section.
An equivalent result was derived in~\autocite{kobayashi_isocurvature_2013} using the $\delta N$ formalism.
Here we provide a more straightforward derivation and extend the use of the formula to more general potentials.

To evaluate expression \eqref{eq:Isocurvature_spectrum}, we will use the fact that at $t_\text{CMB}$ the field should already be oscillating harmonically, as observations require it to behave as cold dark matter already by the time of matter-radiation equality.
As we are already well within the adiabatic regime, the anharmonicity function approach will work well to describe the evolution of the energy density, which means that we can use equation \eqref{eq:anharm_energy_density}.
As fluctuations are small, we can work to linear order in $\sigma_\phi$ to find\footnote{Here we implicitly assume that the fluctuations are still superhorizon when the adiabatic regime is reached. This is indeed the case for all the large scale modes of cosmological interest, like the ones probed by the CMB.}
\begin{equation}\label{eq:isocurvature_anharmonic}
\begin{aligned}
\Delta_\phi^2 &= \Eval{\Braket{ \left( \frac{\delta\rho_\phi}{\rho_\phi}\right) ^2 }}{t_\text{CMB}}{} = \left( \Eval{\frac{\partial \log \rho_\phi(t_\mathrm{CMB})}{\partial \log \phi}}{\phi_0}{}\right)^2 \Braket{ \left( \frac{\delta\phi_0}{\phi_0}\right) ^2 }\\
&=4 \frac{\sigma_\phi^2}{\phi_0^2} \left( 1 + \frac{1}{2} \Eval{ \frac{\d \log f(\theta)}{\d \log \theta}}{\theta_0}{} \right)^2 \\
&= 4\gamma^2 \frac{H_I^2}{4\pi^2 f_a^2 \theta_0^2} \left( 1 + \frac{1}{2} \Eval{ \frac{\d \log f(\theta)}{\d \log \theta}}{\theta_0}{} \right)^2.
\end{aligned}
\end{equation}
Note that even if this quantity is evaluated at $t_\text{CMB}$, it directly depends only on the initial misalignment angle and the statistics of its fluctuations at inflation.
All the information about the later evolution is encoded in the anharmonicity function.

We will now apply the formula \eqref{eq:isocurvature_anharmonic} to both the case of the canonical ALP with a cosine potential and to our non-canonical model, and compare the results with the harmonic approximation.

\begin{figure}[t!]
\centering
   \begin{subfigure}[b]{1\textwidth}
      \includegraphics[width=1\linewidth]{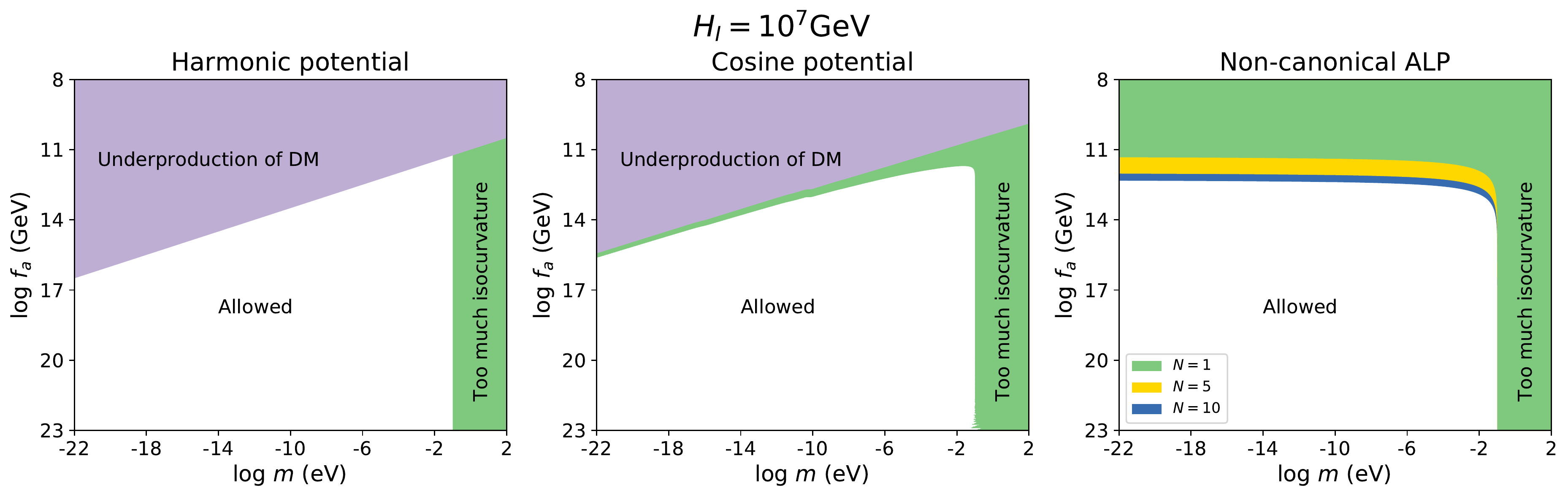}
      %\caption{}
      \label{fig:isocurvature_H7} 
   \end{subfigure}\\[-3ex]

   \begin{subfigure}[b]{1\textwidth}
      \includegraphics[width=1\linewidth]{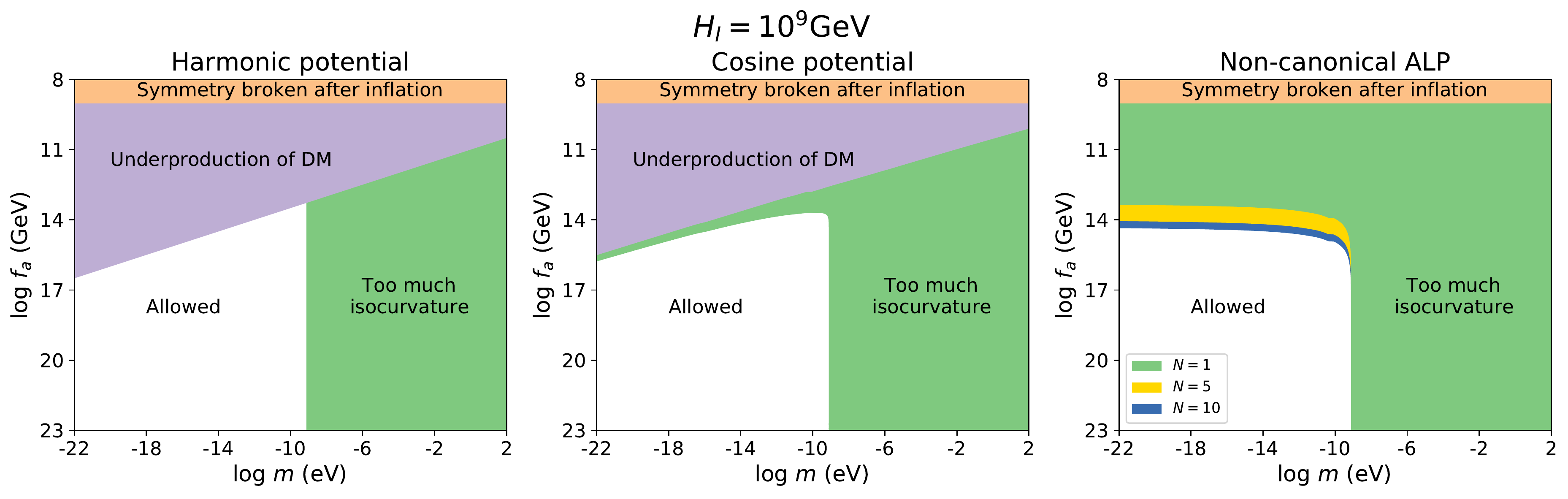}
      %\caption{}
      \label{fig:isocurvature_H9}
   \end{subfigure}\\[-3ex]

   \begin{subfigure}[b]{1\textwidth}
      \includegraphics[width=1\linewidth]{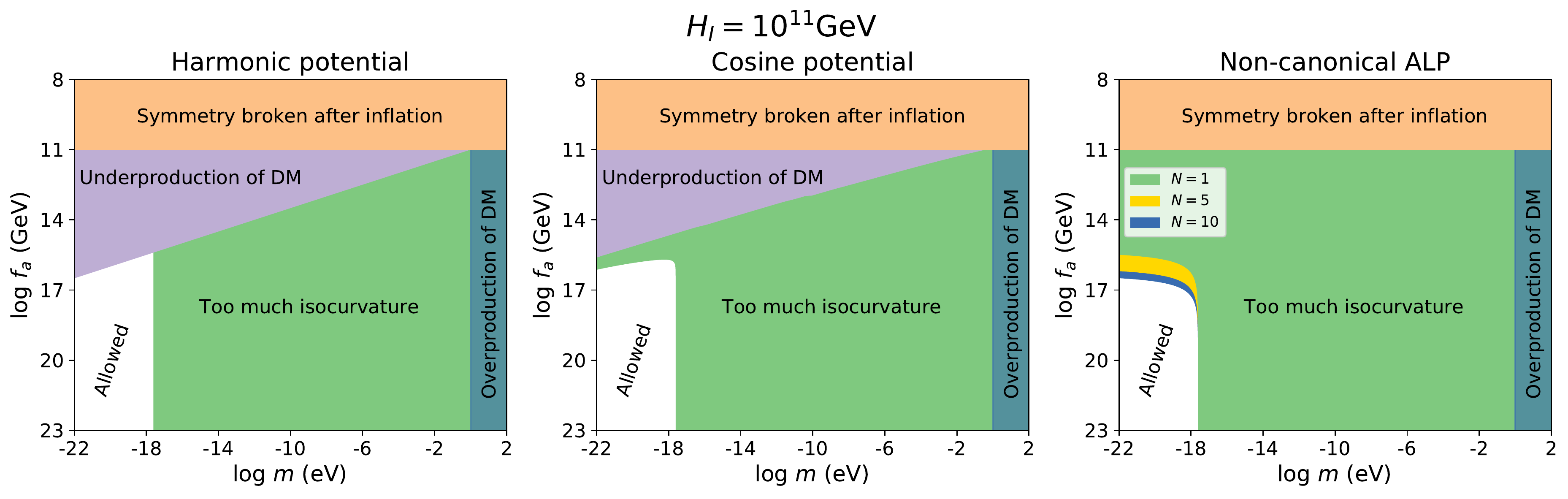}
      %\caption{}
      \label{fig:isocurvature_H12}
   \end{subfigure}\\[-3ex]
   \caption{Isocurvature limits arising in the three different models studied: the harmonic potential (left), the canonical ALP (centre) and the non-canonical one (right). From top to bottom, we plot the limits in the $(m,f_a)$ parameter space for different values of the energy scale of inflation $H_I$. Note that a higher $H_I$ puts stronger bounds on ALP models. In fact, $H_I \gtrsim 10^{12}$ GeV rules out the complete parameter space, whereas for $H_I \lesssim 10^6$ GeV, the limits are very weak.}
   \label{fig:isocurvature}
\end{figure}

For the harmonic case, where $f(\theta_0) = 1$, we have the usual expression
\begin{equation}
\Delta^2_\phi = \gamma^2 \frac{H_I^2}{\pi^2 f_a^2 \theta_0^2}.
\end{equation}
The constraints that one finds, for different values of the energy scale of inflation, are presented in~\figref{fig:isocurvature}.
The harmonic case in particular corresponds to the first column of plots.

Of course, the harmonic case can only be an approximation valid for small $\theta$, as ALP models should preserve the shift symmetry $\theta + 2\pi$.
Among the potentials that satisfy this condition, the most commonly used is $V(\theta) = m^2(1-\cos\theta)$.
The anharmonicity function that appears in this case was studied in~\autocite{turner_cosmic_1986, lyth_axions_1992, strobl_anharmonic_1994, bae_update_2008}.
After comparing with numerical simulations, we have decided to use a slightly different version of it, proposed in~\autocite{diez-tejedor_cosmological_2017} and which provides a better fit to the numerical data,
\begin{equation}\label{eq:cosine_AF}
f(\theta_0) = \left[ \log \left( \frac{e}{1-\left( \theta_0 / \pi\right)^4}\right)\right]^{3/2} .
\end{equation}
With this, it is easy to arrive to the following expression for the isocurvature perturbations,
\begin{equation}
\Delta^2_\phi = \gamma^2 \frac{H_I^2}{\pi^2 f_a^2 \theta_0^2}\left( 1 + \frac{3}{f(\theta_0)^{2/3}} \cdot \frac{1}{(\pi / \theta)^4 -1} \right)^2.
\end{equation}
Note that this reduces to the harmonic result for small $\theta_0$.
However, for angles close to $\pi$, the isocurvature perturbations are greatly enhanced.
As expected, this function diverges at $\theta_0=\pi$, but as we have noted before, this limit is unattainable because of the fluctuations in the field.
In~\figref{fig:isocurvature}, we can see that the limits we can put on the parameter space are a bit stronger than in the harmonic case, in particular for low values of $m$ and $f_a$, which correspond to large values of the initial misalignment angle.

Finally, we turn to the non-canonical case.
The main difference with the canonical ALP, aside from the shape of the potential, is that here we are dealing with an unbounded field range.
As the potential is asymptotically flat, it is always possible to enhance the production of ALPs by choosing a larger initial misalignment angle, as we saw in~\secref{sec:Cosmological_evolution}.
This means that this model can always evade the limits related with to underproduction of dark matter.
Using the anharmonicity function that we derived in the previous section, we find that the isocurvature power spectrum generated in this scenario is
\begin{equation}
\Delta^2_\phi = \gamma^2 \frac{H_I^2}{\pi^2 f_a^2 \theta_0^2}\left( 1 + \frac{1}{2} b N \theta_0 \right)^2,
\end{equation}
where $b=0.56$.
Again, this reduces to the harmonic case for small $\theta$.
The last column of plots in~\figref{fig:isocurvature} illustrate the limits that arise from the Planck data.
Note that in the harmonic and canonical model featuring a compact field range a strong restriction on the parameter space is given by the requirement to produce enough dark matter (the limits arising from this condition are shaded in purple in~\figref{fig:isocurvature}).
As we have already argued, this limit is not present in our non-canonical setup, which features an unbounded field range.
As a consequence, this model opens up a large region of parameter space, corresponding to low masses and decay constants, that was disfavoured until now.

Finally let us remark that, as is well known, high scale inflation strongly constraints ALP models due to the generation of large isocurvature perturbations, which are not seen in the CMB.
The tensor to scalar ratio $r$ is strongly correlated with a high scale of inflation, so a detection of primordial gravitational waves would put a strong constraint on all axion and ALP dark matter models, including ours.
Future experiments~\autocite{matsumura_mission_2014,kogut_primordial_2011,abazajian_cmb-s4_2016} are expected to increase the sensitivity in measuring $r$ and thus the energy scale of inflation.

%%%%%%%%%%%%%%%%%%%%%%%%%%%%%%%%%%%%%%%%%%%%%%%%%%%%%%%%%
\section{Coupling to QCD: Temperature dependent mass}\label{sec:QCD}

So far, we have not assumed a coupling of the  ALP to any other field.
In what follows, we will allow for a coupling to gluons via a term $\theta G \tilde{G}$.
We will study two distinct cases.
First, we contemplate the possibility of having a non-canonical kinetic term in an otherwise QCD-axion model.
Then, we add an extra term to the Lagrangian which, as we will see, allows us to construct a model of light ALPs that enjoys relatively strong gluon couplings.

\subsection{The QCD axion}

In this section we will focus on the QCD axion as introduced by Peccei and Quinn as a solution to the strong CP problem in quantum chromodynamics (QCD)~\autocite{peccei_cp_1977,weinberg_new_1978,wilczek_problem_1978}.

The Lagrangian for the canonically normalised axion field is now
\begin{equation}\label{eq:QCD_lagrangian}
{\cal L}_\phi = \frac{1}{2} \partial^\mu \phi \partial_\mu \phi - \Lambda_{\text{QCD}}^4 \left( 1-\cos \frac{\phi}{f_a}\right),
\end{equation}
and as usual we can define the angle $\theta = \phi/f_a$, so that $\theta\in \left( -\pi,\pi \right]$. 
For our modification with a non-canonically normalised field, we have
\begin{equation}\label{eq:QCD_noncan_lagrangian}
{\cal L}_{\phi} = \frac{1}{2 \cos ^2 \left( N \frac{\phi}{f_a} \right)} \partial^\mu \phi  \partial_\mu \phi  - \Lambda_{\text{QCD}}^4 \left( 1-\cos \frac{\phi }{f_a}\right),
\end{equation}
and after we perform a field redefinition to have it canonically normalised, we find the Lagrangian
\begin{equation}\label{eq:QCD_noncan_lagrangian_norm}
{\cal L}_\varphi = \frac{1}{2} \partial^\mu \varphi \partial_\mu \varphi 
- \Lambda_{\text{QCD}}^4 \left[ 1-\cos \left( \frac{2}{N} \arctan \left( \tanh \frac{N\varphi}{2f_a} \right) \right) \right].
\end{equation}
There is just one difference that makes the QCD axion case particular, and it is that here the energy scale appearing in the potential is fixed by QCD to be~\autocite{di_cortona_qcd_2016}
\begin{equation}\label{eq:QCD_energy_scale}
\Lambda_{\text{QCD}} = f_\pi m_\pi \frac{\sqrt{m_u m_d}}{m_u + m_d} \simeq 76\ \text{MeV}.
\end{equation}
It is easy to see that the mass of the axion, $m_a$, is given by $f_am_a = \Lambda_{\text{QCD}}^2$.
It is important to note that the numerical value quoted above is only valid at zero (or very low) temperatures.
Indeed, the axion potential is affected by finite temperature effects, such that the mass of the axion varies with temperature.
At low temperatures below the QCD critical temperature $T_\mathrm{crit}\sim 160-170\ \text{MeV}$, the mass remains roughly constant\footnote{The small temperature dependence can be computed using chiral perturbation theory as in~\autocite{di_cortona_qcd_2016}}.
That said, much of the dynamics that is of interest to us will happen in the early universe, at temperatures close or above $T_\mathrm{crit}$. There are different ways to compute the temperature dependence of the axion mass~\autocite{turner_cosmic_1986, bae_update_2008, wantz_axion_2010, wantz_topological_2010, di_cortona_qcd_2016, borsanyi_axion_2016}.
The function that controls the temperature dependence of the axion mass is the topological susceptibility $\chi(T)$, which is usually parametrised as a power law:
\begin{equation}\label{eq:axion_mass_T}
m_a^2(T) = \frac{\chi (T)}{f_a^2}, \quad \text{where}\quad \chi(T) \simeq \chi_0 \left( \frac{T}{T_{\mathrm{crit}}}\right)^{2\alpha}.
\end{equation}
Here we will use $2\alpha = -7.1$ and $\chi_0 = 0.11$, from recent lattice computations~\autocite{borsanyi_axion_2016} that are consistent with the instanton values up to an overall normalisation factor.

We see that the main effect is that the mass of the axion is approximately constant until $T_\mathrm{crit}$, and then it drops as a power law, so that the axion is essentially massless at high temperatures.
The most important implication of the temperature dependent mass is that a smaller mass at early times can delay the start of the oscillations of the field, which in turn results in a higher energy density of axionic dark matter today.
This happens both for the canonical and non-canonical axion models.

\subsection{Anharmonicity function and isocurvature perturbations revisited: Temperature dependence}

We have seen that coupling ALPs to QCD through $\phi G\tilde{G}$ results in a temperature-dependent mass for the ALP, both in the canonical and non-canonical setup.
This of course has an impact on its cosmological evolution, which can be of importance in computing observables such as the isocurvature perturbations that we discussed in~\secref{sec:Isocurvature_perturbations}.
To account for this effect, we will modify the anharmonicity function formalism that we introduced in~\secref{sec:numerics} to incorporate the temperature dependence.
That is, we want to compute
\begin{equation}
F_T(\theta_0,f_a) \equiv \frac{\rho^{\mathrm{anh}}_T}{\rho^{\mathrm{harm}}},
\end{equation}
evaluated at a point in time late enough so that the anharmonic and temperature-dependent axion field has already entered the adiabatic regime.

\begin{figure}[t!]
\centering
      \includegraphics[width=0.6\linewidth]{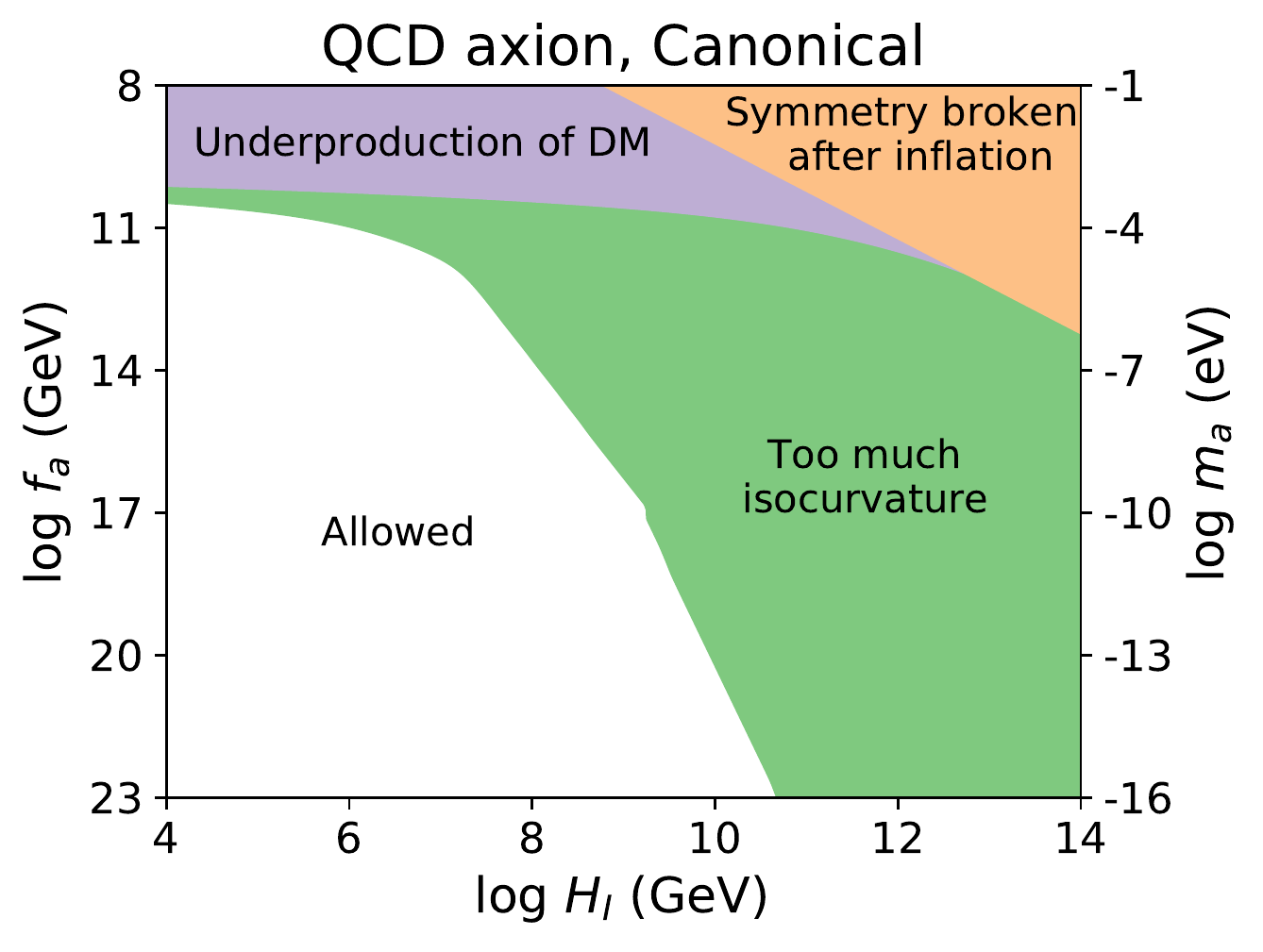}
      \caption{Isocurvature constraints on the axion scale $f_{a}$ as a function of the inflation scale $H_{I}$ for the QCD axion with potential \eqref{eq:QCD_lagrangian}. Both the anharmonicities of the potential and the temperature dependence of the mass are taken into account through the anharmonicity function defined in \eqref{eq:temperature_anharmonicity_function}. Our results differ slightly from the ones obtained in~\autocite{visinelli_dark_2009} and~\autocite{kobayashi_isocurvature_2013} due to the fact that we are using the more recent data from the Planck satellite and a different anharmonicity function.}
      \label{fig:QCD_isocurvature_cosine}
\end{figure}

For definiteness, we will use the following expression for the axion mass,
\begin{equation}\label{eq:QCD_axion_mass}
m_a(T) = \begin{cases}
m_a \left( \frac{T}{T_\mathrm{crit}} \right)^\alpha  &\mathrm{if}\ T\geq T_\mathrm{crit} ,\\
m_a  &\mathrm{if}\ T\leq T_\mathrm{crit} .\\
\end{cases}
\end{equation}
First of all, we note that this temperature dependence will only have an effect if the field starts oscillating before the QCD critical temperature $T_\mathrm{crit}$.
In the harmonic limit, this means that if the mass is smaller than $m_a^*$, defined by $3H(T_\mathrm{crit})=m_a^*$, the field will have acquired its late-time mass by the time it starts oscillating.
Thus, the later evolution of the field will be insensitive to the temperature effects that happened earlier on.
In terms of decay constants, this sets a distinct scale
\begin{equation}
f_a^* \simeq 8.7\cdot 10^{16}\ \mathrm{GeV}.
\end{equation}

If we take into account the anharmonicities of the potential, it might happen that the start of the oscillations is delayed until after $T_\mathrm{crit}$, even if $f_a<f_a^*$.
The condition to be in this regime is that the initial misalignment angle $\theta_0$ is larger than some value $\theta_0^* (f_a)$.
This value is given for a general anharmonicity function\footnote{Note the difference between $f(\theta_0)$, which is the anharmonicity function presented in~\secref{sec:numerics} and induced purely by the shape of the potential, and $F_T(\theta_0,f_a)$, which also includes the effects of the temperature-dependent mass.} by
\begin{equation}
\left( \frac{f_a^*}{f_a} \right)^{3/2} = f(\theta_0^*).
\end{equation}
For the case of a canonical axion with a cosine potential like in \eqref{eq:QCD_lagrangian}, we find
\begin{equation}
\theta_0^* (f_a) \simeq \pi \left[ 1 - \exp \left( 1 - \frac{f_a^*}{f_a} \right) \right]^{1/4},
\end{equation}
whereas in the non-canonical case \eqref{eq:QCD_noncan_lagrangian_norm}, we find
\begin{equation}
\psi_0^* (f_a,N) \simeq \frac{3}{2bN} \log \left( \frac{f_a^*}{f_a} \right).
\end{equation}

\begin{figure}[t!]
\centering
      \includegraphics[width=0.6\linewidth]{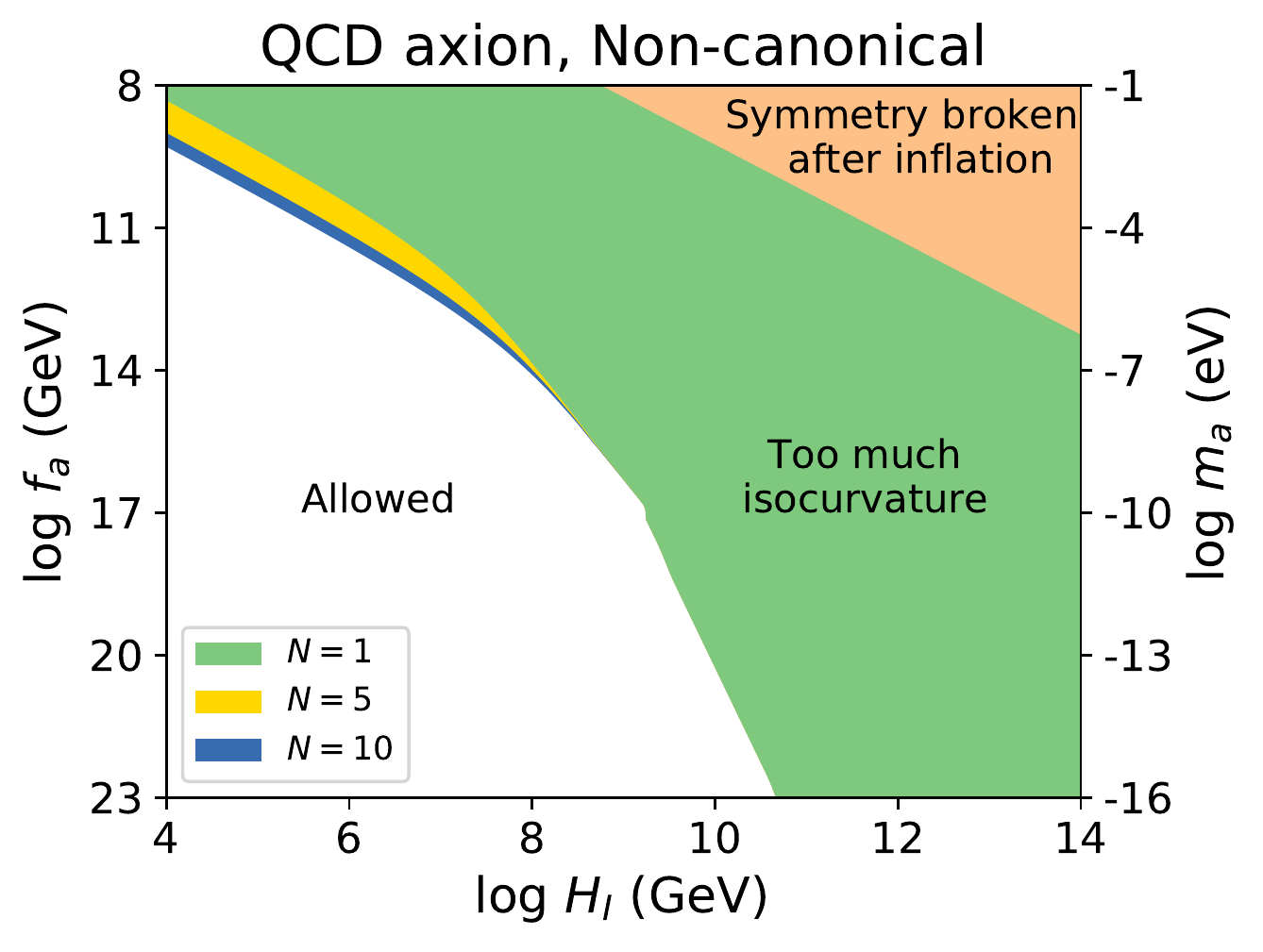}
      \caption{Isocurvature constraints on the axion scale $f_{a}$ as a function of the inflation scale $H_{I}$ for the QCD axion with a non-canonical kinetic term \eqref{eq:QCD_noncan_lagrangian}. Both the anharmonicities of the potential and the temperature dependence of the mass are taken into account through the anharmonicity function defined in \eqref{eq:temperature_anharmonicity_function}.}
      \label{fig:QCD_isocurvature_noncan}
\end{figure}

For any set of decay constants and initial misalignment angles that satisfy $f_a<f_a^*$ and $\theta_0<\theta_0^*$, we compute $F_T$ for a generic anharmonic potential, finding
\begin{equation}
F_T(\theta_0,f_a) \simeq \left( \frac{f_a^*}{f_a} \right)^{\frac{\alpha}{2(2-\alpha)}}\cdot\left( f (\theta_0) \right)^{\frac{2(3-\alpha)}{3(2-\alpha)}}.
\end{equation}
The details of the derivation of this result are given in Appendix~\secref{sec:Appendix2}.
Here we see that the result depends critically on the exponent of the temperature-dependence of the axion mass at high temperatures above the QCD critical temperature.
To sum up, we can write the full temperature-dependent anharmonicity function as follows,
\begin{equation}\label{eq:temperature_anharmonicity_function}
F_T(\theta_0,f_a) = 
\begin{cases}
f(\theta_0) &\mathrm{if}\ f_a > f_a^*,\\
f(\theta_0) &\mathrm{if}\ f_a < f_a^*\ \mathrm{and}\ \theta_0 > \theta_0^*,\\
\left( \frac{f_a^*}{f_a} \right)^{\frac{\alpha}{2(2-\alpha)}}\cdot\left( f (\theta_0) \right)^{\frac{2(3-\alpha)}{3(2-\alpha)}}\quad &\mathrm{if}\ f_a < f_a^*\ \mathrm{and}\ \theta_0 < \theta_0^*.
\end{cases}
\end{equation}
With this, we can use the same approach as in~\secref{sec:Isocurvature_perturbations_anharmonic} to compute the isocurvature perturbations, this time using the temperature-dependent anharmonicity function,
\begin{equation}\label{eq:isocurvature_anharmonic}
\Delta_\phi^2 = 4\gamma^2 \frac{H_I^2}{4\pi^2 f_a^2 \theta_0^2} \left( 1 + \frac{1}{2} \Eval{ \frac{\d \log F_T(\theta)}{\d \log \theta}}{\theta_0}{} \right)^2.
\end{equation}

We apply this formula for both the canonical QCD axion and for our non-canonical model, and obtain the results presented in \figref{fig:QCD_isocurvature_cosine} and \figref{fig:QCD_isocurvature_noncan}, respectively.
For the canonical QCD axion, our results are an update from the ones obtained in~\autocite{visinelli_dark_2009} and~\autocite{kobayashi_isocurvature_2013}, as we are using the more recent data from the Planck satellite and a better fitting anharmonicity function.

\subsection{ALPs coupled to QCD with $f_{a} m << \Lambda_{\rm QCD}^2$}

Let us now study the possibility of an ALP having a coupling to the $G\tilde{G}$ term while satisfying $f_a m << \Lambda_{\rm QCD}^2$.
This is an interesting region of the parameter space, as ALPs that satisfy these conditions may be found by looking for an oscillating nucleon or atomic electric dipole moment.
There exist a number of proposed laboratory searches focusing on this direction~\autocite{graham_new_2013,graham_axion_2011,budker_cosmic_2014,hexenia_2017} .

However, we have seen that coupling the ALP to QCD via a term proportional to $G\tilde{G}$ induces an irreducible contribution to the mass, given by \eqref{eq:QCD_energy_scale}.
Explicitly this contributes
\begin{equation}\label{eq:fatoma}
m^{2}_a (T=0) \simeq \left(5.7\times 10^{-5} {\rm eV }\left( \frac{10^{11}{\rm GeV}}{f_a} \right)\right)^2,
\end{equation}
to the square of the axion mass as given in~\autocite{di_cortona_qcd_2016}.
This contribution will also have a temperature dependence as described by \eqref{eq:axion_mass_T}.

A priori, this irreducible contribution to the axion mass seems irreconcilable with the condition $f_a m << \Lambda_{\rm QCD}^2$~\autocite{blum_constraining_2014}.
The only known way of circumventing this caveat is to precisely cancel this contribution with an additional, fine-tuned term in the Lagrangian.
Acknowledging the flaws of this \textit{ad hoc} approach, we follow it and study the phenomenology of such models when allowing for a non-canonical kinetic term.

At the level of the Lagrangian, we add an extra term to the potential so that it becomes
\begin{equation}
V(\phi) = \Lambda_{\rm QCD}^4\left( 1-\cos \frac{\phi}{f_a} \right) - \Lambda_0^4\left(1- \cos \left(\frac{\phi}{n f_a} + \alpha \right) \right).
\end{equation}

In principle there can exist a phase difference between both contributions.
For our purposes, it will be necessary to require that this phase difference vanishes, so we will take $\alpha=0$.
This can be viewed as equivalent to asking for a separate a solution to the strong CP problem.
In principle any integer $n$ is possible but for simplicity we will limit ourselves to the $n=1$ case.
In the small $\phi$ limit, this potential induces a mass for the ALP
\begin{equation}
m^2 = m_a^2(T) - m_0^2,
\end{equation}
where $m_0f_a = \Lambda_0^2$ and recall that $m_a$ is completely fixed by $f_a$ as in equation \eqref{eq:fatoma}. 
It is then possible to choose $m_0$ so that we get any zero-temperature mass for the ALP, i.e. we can set $m_0^2 = m_a^2(T=0) - m^2$.
We are interested in the $m^2 \ll m_a^2(T=0)$ regime.
The full mass can then be expressed as 
\begin{equation}
m^2 (T) = m_a^2(T) - m_a^2(0) + m^2.
\end{equation}
Because at early times the QCD contribution is strongly suppressed, in that regime we have $m^2 (T) < 0$.
We will use the following simplified expression for the temperature dependent mass of the ALP
\begin{equation}\label{eq:temp_mass}
m^2 (T) = \begin{cases}
m^2 & \quad{\rm for}\quad T<T_{\rm crit} \\
-m_a^2(0) & \quad{\rm for}\quad T>T_{\rm crit}
\end{cases}
\end{equation}
Note that the negative mass does not indicate an unstable potential but only that $\phi=0$ is not the minimum at that time.

\subsubsection{Canonical case}
\begin{figure}[t!]
\centering
\includegraphics[width=0.5\linewidth]{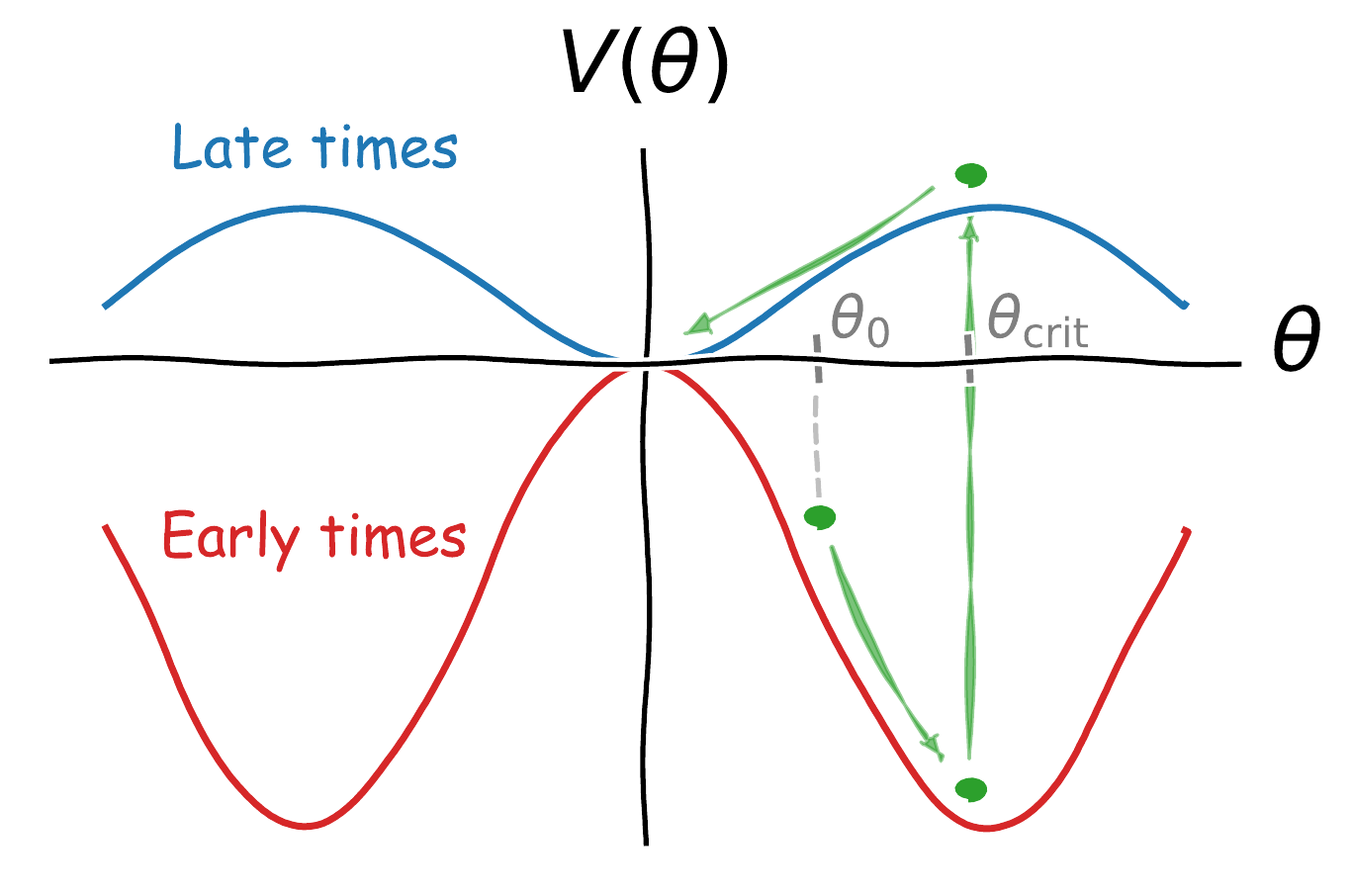}
\caption{Cartoon explaining the evolution of the field.
The red and blue lines represent the potential before and after the QCD phase transition, respectively.
The green dots and arrows represent the evolution of the field. (The oscillations are not drawn explicitly in order to simplify the figure.)
The initial misalignement angle $\theta_0$ and the value of the field at the QCD phase transition, $\theta_{\mathrm{crit}}$, are depicted.}
\label{fig:N0_sketch} 
\end{figure}

\begin{figure}[t!]
\centering
\includegraphics[width=0.7\linewidth]{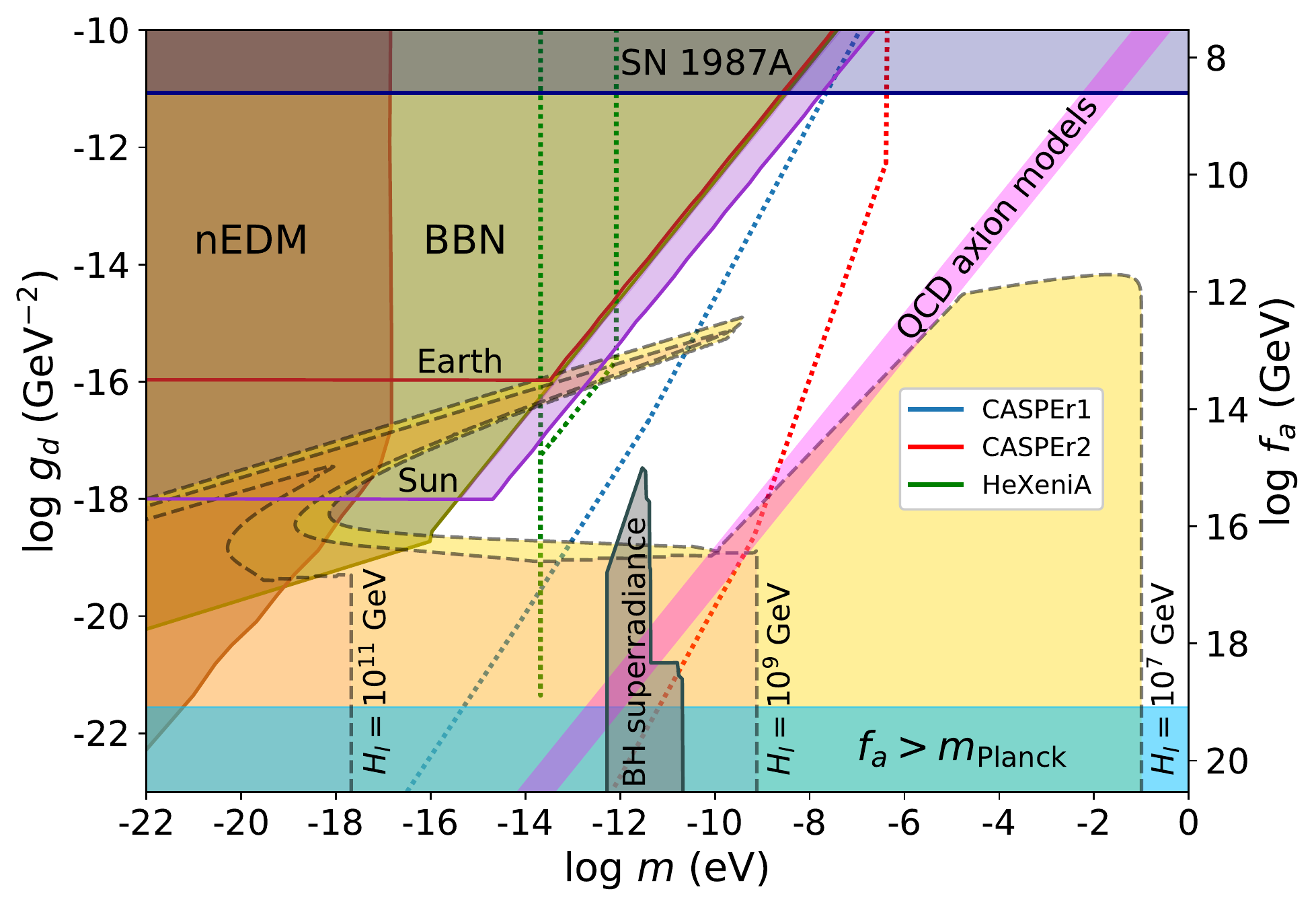}
\caption{Parameter space of the model defined by the potential given in \eqref{eq:temp_dep_N0}.
We present the isocurvature constraints for different values of $H_I$, ranging from $10^6$ to $10^{11}$ GeV.
A higher scale of inflation restricts the model to lie in the respective coloured areas.
For the purpose of visualisation we have continuously connected the solution in the two different regimes that we have considered, i.e. to the left and to the right of the QCD axion band.
All the other limits presented in~\figref{fig:ALP_parameter_space} are also applicable in this scenario, as they only depend on the dynamics of the field after the QCD phase transition.}
\label{fig:N0_parameter_space} 
\end{figure}

As a first step, we implement the mass subtraction and the resulting temperature dependence in an ALP model with a canonically normalised scalar field with potential given by
\begin{equation}\label{eq:temp_dep_N0}
V(\phi) = f_a^2\ m^2 (T) \left( 1-\cos \frac{\phi}{f_a} \right),
\end{equation}
with \(m(T)\) defined in \eqref{eq:temp_mass}.
The most relevant feature of this scenario is that before the QCD phase transition, the potential is minimised at $\theta=\pi$ rather than at $\theta=0$.
Accordingly, at early times the field evolves towards its minimum at $\pi$, around which it will oscillate with damped amplitude.
Then, after the QCD phase transition, the potential rapidly acquires its late-time shape, with a minimum at the origin.
The field thus oscillates around its CP-conserving value $\theta=0$ at late times.
The main role of the first set of oscillations is to set the initial condition for the second one to be close to $\pi$.
We refer to \figref{fig:N0_sketch} for a cartoon explaining this evolution.
It should be noted that this discussion is only valid if $f_am\ll \Lambda^{2}_{\mathrm{QCD}}$, that is, if we lie to the left of the QCD axion band in \figref{fig:N0_parameter_space}.
In the other limit, \textit{ie} $f_am\gg \Lambda^{2}_{\mathrm{QCD}}$, the contribution of the QCD mass is negligible and we recover the usual constant mass ALP scenario.

Let us now be a bit more quantitative.
Initially, $H$ is large and the field is stuck at its initial value $\theta_0$.
Then, as long as the early-time mass $m_a(0)$ overcomes the Hubble friction before the QCD phase transition, the field will oscillate around $\pi$.
The condition for this to happen is roughly $f_a\gtrsim10^{17}$ GeV, but this value can be modified by the anharmonicities depending on the initial misalignment.
These oscillations continue until the temperature decreases to $T_{\rm crit}$, at which time the amplitude is approximately given by
\begin{equation}
\left( \pi-\theta_{\mathrm{crit}} \right) \simeq \left( \pi-\theta_0 \right)\left( \frac{\mathcal{F}(T_{\rm crit})}{\mathcal{F}(T_1)} \right)^{1/2}\left( \frac{f_a}{2\cdot 10^{17}\ \mathrm{GeV}} \right)^{3/4} f^{1/2}(\theta_0).
\end{equation}
Here, the anharmonicity function is given by \eqref{eq:cosine_AF} and $T_1$ is defined by $3H(T_1)=m_a(0)$.
The value of $\theta_{\mathrm{crit}}$ gives the initial condition for the oscillations that happen after the QCD phase transition, now around $\theta=0$ and with frequency given by the late-time mass $m$.
Typically, $\theta_{\mathrm{crit}}$ is very close to $\pi$ so the anharmonicites of the potential will play a key role.
Taking this into account, we can compute the energy density of the oscillating scalar field as
\begin{equation}\label{eq:N0_energy_density}
\rho\simeq 0.17 \frac{\mathrm{keV}}{\mathrm{cm}^3}\ \mathcal{F}(T_2)\ \sqrt{\frac{m}{\mathrm{eV}}}\left(\frac{f_a}{10^{11}\ \mathrm{GeV}}\right)^2 \ \theta^2_\mathrm{crit}\ f(\theta_{\mathrm{crit}}),
\end{equation}
where $3H(T_2)=m$.

We can then determine in what region of parameter space the right dark matter abundance can be generated with an initial misalignment angle $\theta_0$  of order $\orderof (1)$.
It is possible to either enhance or suppress the energy density given in \eqref{eq:N0_energy_density} by tuning the initial misalignment angle closer to zero or $\pi$.
However, due to equation \eqref{eq:isocurvature_anharmonic}, there is an enhancement of the isocurvature perturbations each time the field gets close to a maximum of the potential, where the anharmonicity function becomes large.
Because of this, the available tuning of the initial misalignment angle is very limited in this scenario due to the stringent constraints on isocurvature fluctuations.
\figref{fig:N0_parameter_space} shows how the allowed parameter space shrinks for larger values of the Hubble scale of inflation.
Despite the strong isocurvature constraints, we can see that this scenario populates some unexplored regions of parameter space to the left of the QCD axion line that could be probed by upcoming experiments looking for ALPs.

\subsubsection{Non-canonical case}
We now want to implement the temperature dependent potential~\eqref{eq:temp_dep_N0}
 in our non-canonical ALP scenario.
In terms of the cosmological evolution of the field, this is effectively done by writing
\begin{equation}
V(\varphi) = f_a^2 m^2 (T) \left[ 1 - \cos \left( \frac{2}{N} \arctan \left( \tanh \frac{N\varphi}{2f_a} \right) \right) \right].
\end{equation}
This is the same potential as we had before, except that for high temperatures $T>T_{\rm crit}$ the mass squared will be negative and will be a function only of $f_a$, as given in \eqref{eq:temp_mass}.
This tells us that, depending on the value of the parameters $m$ and $f_a$, we will have two very different behaviours, which qualitatively can be understood as follows.

First, if the field does not start rolling until after the QCD phase transition, then all the dynamics and the observables will not be affected at all by the features of the potential at high temperatures.
This is because there is no evolution while the field is frozen by Hubble friction.
Only after it has acquired its late time mass $m$ does it start rolling, and thus the cosmological evolution is exactly as we computed in~\secref{sec:Cosmological_evolution}.
However, the key difference is that now the ALP is coupled to $G\tilde{G}$, so it may be tested by observables and experiments that exploit this coupling.

The other option is, of course, that the field starts rolling before the QCD phase transition.
Then the dynamics can depend strongly on the initial conditions and is rather complicated.
However, we will see that this scenario leads to an overproduction of ALPs whose energy density exceeds the observed CDM one.
As we are only interested in ALPS as dark matter candidates, the second scenario is not interesting for us and we just need to focus on the first one.

Let us now be more quantitative and compute what region of the parameter space  allows for ALP dark matter with a non-canonical kinetic term and coupled to QCD.
As we have anticipated, this ALP will only be a good dark matter candidate if its evolution is frozen until after the QCD phase transition.
Then, the present ALP energy density will only depend on $f_a$, the present mass $m$ and the initial misalignment angle $\psi_0$.
The latter is given by equation \eqref{eq:anharm_energy_density}, and satisfies
\begin{equation}\label{eq:initial_misalignment_angle}
\psi_0^2\ \me^{bN\psi_0} \simeq \frac{7.26}{{\cal F}(T_1)} \sqrt{\frac{{\rm eV}}{m}} \left( \frac{10^{11}\text{ GeV}}{f_a} \right)^2,
\end{equation}
where $T_1$ is the temperature at which the oscillations start.

\begin{figure}[t!]
\centering
   \begin{subfigure}[b]{0.49\textwidth}
      \includegraphics[width=1\linewidth]{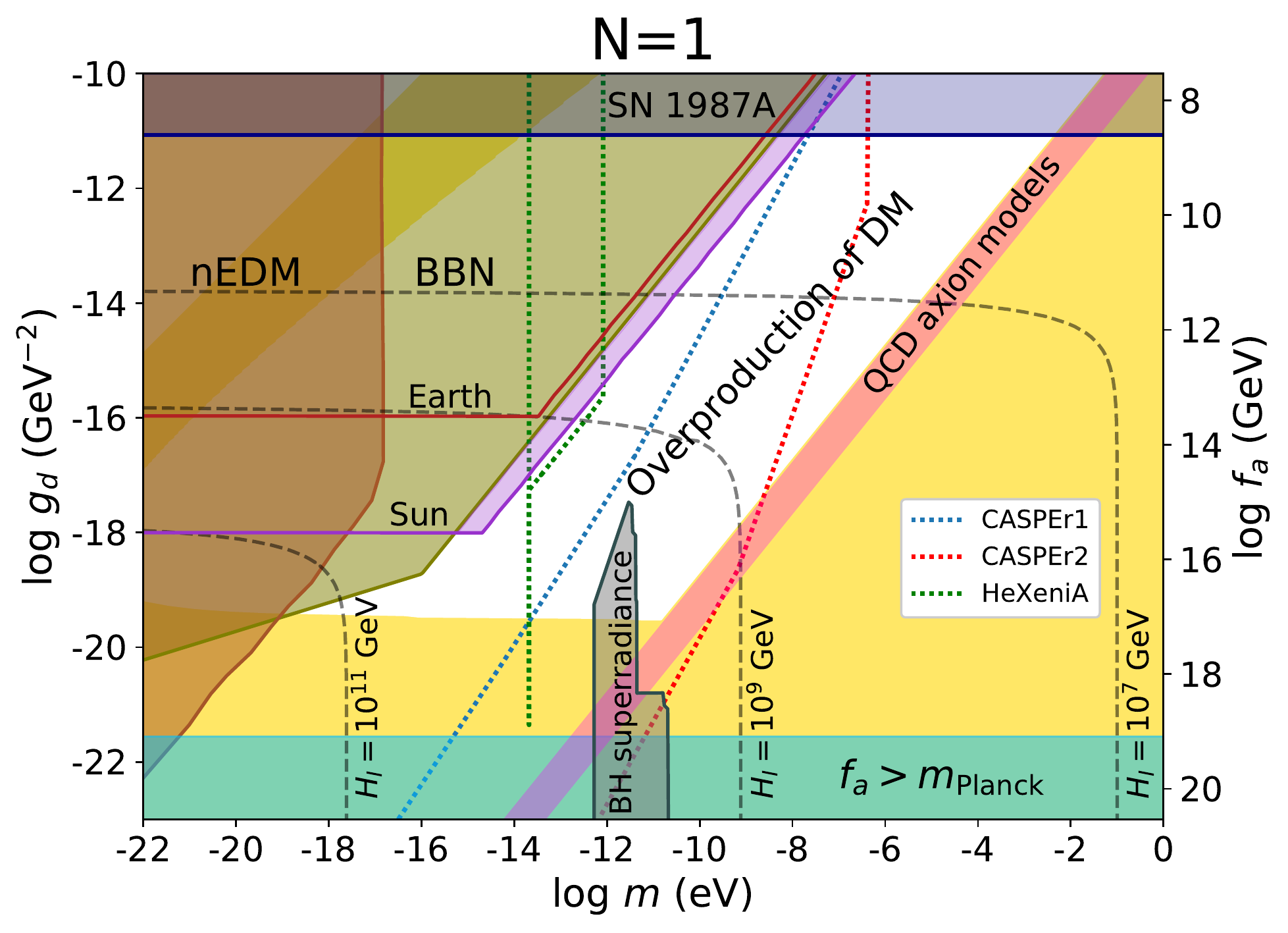}
      %\caption{}
      \label{fig:Noncan_QCD_parameter_space_N1} 
   \end{subfigure}
   \begin{subfigure}[b]{0.49\textwidth}
      \includegraphics[width=1\linewidth]{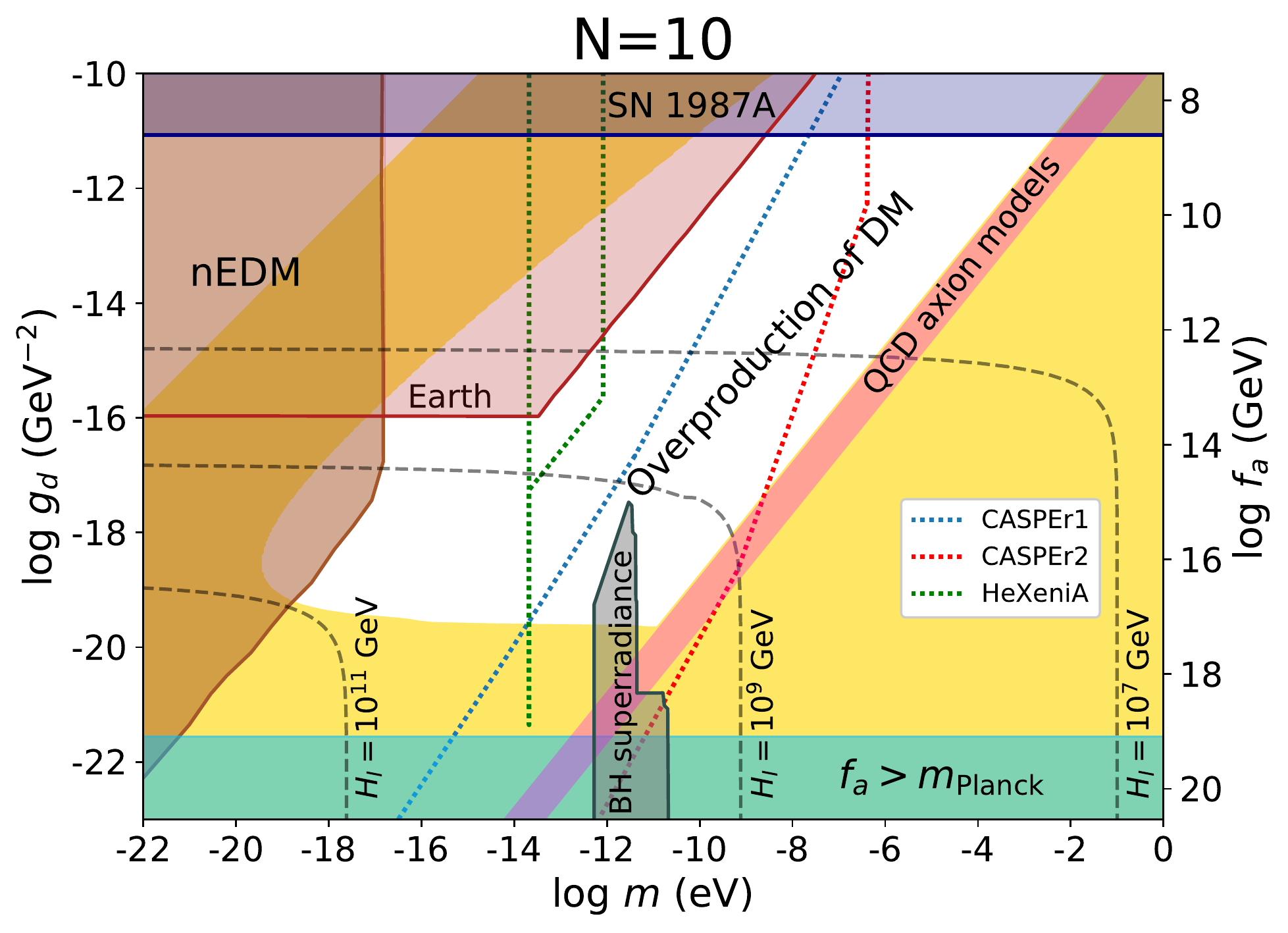}
      %\caption{}
      \label{fig:Noncan_QCD_parameter_space_N10}
   \end{subfigure}\\[-3ex]

   \begin{subfigure}[b]{0.49\textwidth}
      \includegraphics[width=1\linewidth]{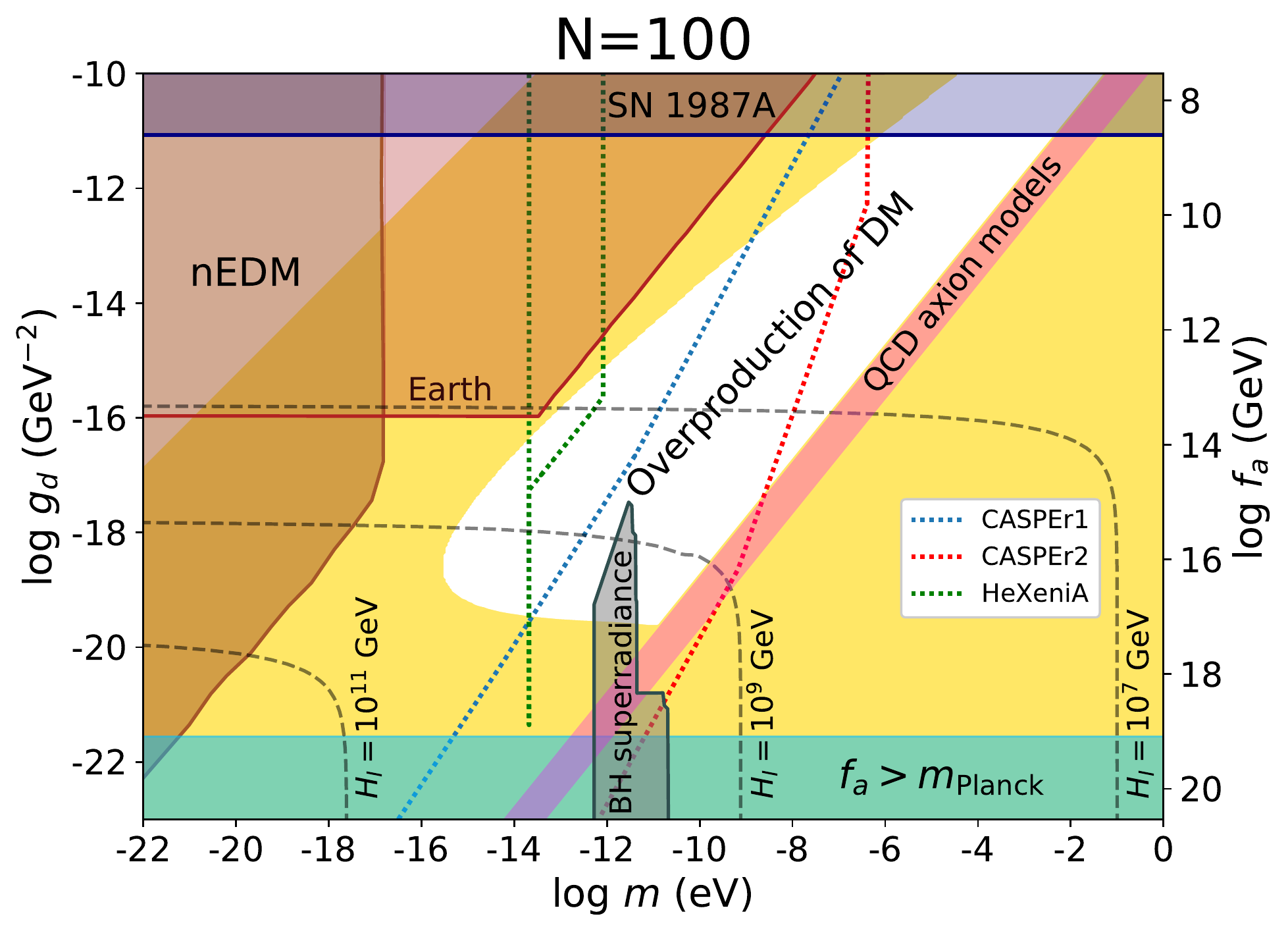}
      %\caption{}
      \label{fig:Noncan_QCD_parameter_space_N100}
   \end{subfigure}
   \begin{subfigure}[b]{0.49\textwidth}
      \includegraphics[width=1\linewidth]{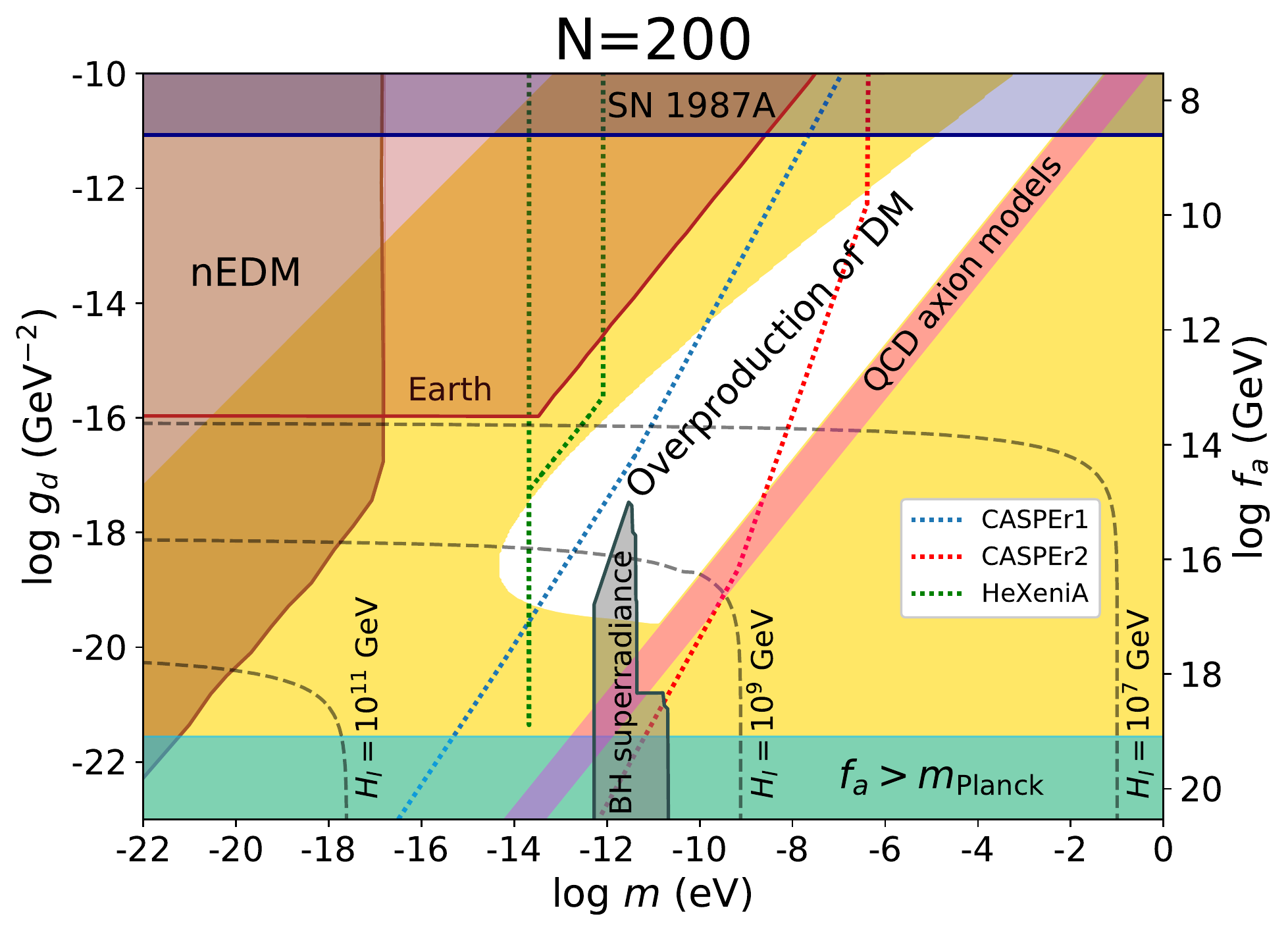}
      %\caption{}
      \label{fig:Noncan_QCD_parameter_space_N200}
   \end{subfigure}\\[-3ex]
\caption{Parameter space for ALPs with a non-canonical kinetic term of the form \eqref{eq:non_canonical_kinetic_term} coupled to QCD via a $G\tilde{G}$ term, along with constraints coming from its cosmological evolution and searches for an oscillating EDM. Each panel represents a different value of the parameter $N$. This scenario can provide the right dark matter density in the yellow shaded region, while the areas excluded by overproduction of dark matter or the condition \eqref{eq:second_inflation_limit} to avoid a second period of inflation are coloured in white. The brown region is excluded by re-analysing data originally intended to search for a static neutron EDM in order to look for an oscillating one~\autocite{abel_search_2017}. The dark green region in the first figure is inconsistent with the production of the observed abundance of light elements during Big Bang Nucleosynthesis~\autocite{blum_constraining_2014}. This limit is effective only for $N<4$ and absent in the other figures. Similarly, the limit from~\autocite{hook_probing_2017} corresponding to the ALP field being sourced at the Sun only applies for small values of $N$, while the Earth one stays valid in all cases. Finally, $f_a$ is (softly) bounded from above by the requirement that it does not exceed the Planck scale, and from below by the supernova limits estimated in~\autocite{raffelt_astrophysical_2008}.}
\label{fig:temperature_BBN_EDM}
\end{figure}

We now need to find what the region of the parameter space is where the field starts oscillating only after the QCD phase transition.
The QCD phase transition happens at a temperature of around $T_{\rm crit}\sim 160\ {\rm MeV}$, which corresponds to a Hubble parameter of $H(T_{\rm crit})\sim 10^{-11}\ {\rm eV}$.
By asking that $3H(T_{\rm crit}) > | V^{\prime\prime}(\psi_0) |^{1/2}$, we get the condition
\begin{equation}\label{eq:QCD_rolling_condition}
H(T_{\rm crit}) > 3.24\times 10^{-6}{\rm eV }\cdot \sqrt{2N\sin\frac{\pi}{2N}} \ {\cal F}(T_1)^{1/(2b)}\ \psi_0^{1/b} \left( \frac{m}{{\rm eV}} \right)^{1/(4b)} \left( \frac{f_a}{10^{11}\text{ GeV}} \right)^{1/b-1}.
\end{equation}
This region is plotted in~\figref{fig:temperature_BBN_EDM}, together with the further cosmological and astrophysical bounds that restrict the parameter space.

Finally we still have to justify our claim that if the field starts oscillating before the QCD phase transition we always get an overproduction of ALPs.
For a given $(m,f_a)$, any initial misalignment angle bigger than the one given by \eqref{eq:initial_misalignment_angle} will lead to an energy density in ALPs greater than the observed dark matter one.
But if the condition \eqref{eq:QCD_rolling_condition} is not satisfied, then the field will start rolling towards bigger $\psi$ values, because $m^2(T)<0$ at high temperatures.
Thus, the effect of the rolling at high temperatures is to drive the field away from the required misalignment angle to give the correct dark matter abundance.
This statement is independent of what misalignment angle we start with, and thus rules out ALPs in the region coloured in white in~\figref{fig:temperature_BBN_EDM} as dark matter candidates\footnote{Such an overproduction could, e.g. be ameliorated in scenarios with two stages of inflation~\autocite{davoudiasl_inflatable_2016,hoof_qcd_2017}.}.

\subsection{Big Bang Nucleosythesis}\label{secbbn}
Aside from a potential over (or under) production there is an additional constraint that rules out large areas of experimentally accessible parameter space. This arises from cosmology, more precisely BBN~\autocite{blum_constraining_2014}. A non-vanishing $\theta$ angle at the time of BBN can spoil the production of light elements such as $^4$He.
This is due to the fact that a non-vanishing $\theta$ angle induces a difference between the mass of the proton and the neutron~\autocite{ubaldi_effects_2010}
\begin{equation}
\delta Q \equiv m_n-m_p = c_+ \frac{m_d^2-m_u^2}{\sqrt{m_u^2+m_d^2+2m_um_d\cos\theta}},
\end{equation}
where $c_+\simeq 2.5$ can be determined by looking at the mass splitting $M_\Theta-M_N$ in the baryon octet~\autocite{crewther_chiral_1979}.
A larger mass splitting means that the freeze-out abundance of neutrons with respect to protons would be lower.
In addition to that, the free neutron decay rate is enhanced, which means that more neutrons decay between freeze-out and nucleosynthesis.
This depletion of neutrons\footnote{There are other effects that play a role, like the change in the deuteron binding energy or the rise in the freeze-out temperature. We have found that the contribution of these effects is smaller than the one considered above, so we neglected them for this analysis.} eventually turns into an underproduction of $^4$He.
Based on the discussion in~\autocite{blum_constraining_2014,stadnik_can_2015}, these effects result in a shift that can be estimated as
\begin{equation}
\frac{\delta Y_p}{Y_p} \equiv \frac{Y_p^0-Y_p(\theta)}{Y_p^0} = \left( 1- \frac{Y_p^0}{2} \right) \left( \frac{\delta \left(n/p\right)_{\mathrm{fr}}}{\left(n/p\right)_{\mathrm{fr}}} + \delta\Gamma_n t_{\mathrm{nuc}} \right) \simeq 0.66 \theta^2.
\end{equation}
Using the values $Y_p^0=0.25$ and $t_{\mathrm{nuc}}=880$ s~\autocite{mukhanov_physical_2005}.
One can now take the conservative limit $|\delta Y_p/Y_p|<10\%$ to see that successful nucleosynthesis requires
\begin{equation}
\theta_\mathrm{BBN} < 0.39.
\end{equation}
In our non-canonical model, the first thing we notice is that the $\theta$ angle is bounded,
\begin{equation}
| \theta | = | \frac{2}{N}\arctan\left( \tanh \frac{N\psi}{2} \right) | \leq \frac{\pi}{2N},
\end{equation}
so the BBN bound is completely avoided if $N>4$.
For smaller N we are in the region of small $\theta$ and the behaviour is approximately that of a canonical ALP. 
Here we use the bound given in~\autocite{blum_constraining_2014}.
The corresponding excluded region is shaded in darker and labeled ``BBN'' in~\figref{fig:temperature_BBN_EDM}.

\section{Conclusions}\label{sec:Conclusions}

The question raised in this paper can be summarised in the following way: Is it possible to have an axion-like particle (ALP) with a non-canonical kinetic term as a phenomenologically viable and interesting dark matter candidate? Our study points towards an affirmative answer.
Using in particular a non-canonical term with singularities similar to those used in $\alpha$-attractor models for inflation we find a significantly enlarged parameter space for dark matter.
In particular, regions with larger couplings -- where canonical ALPs are underproduced -- now become viable, offering interesting possibilities for near future experiments.

For the production via the misalignment mechanism the key feature of the non-canonical kinetic term is that today's ALP energy density is enhanced due to a delay in the start of the oscillations.
This arises because the effective potential is flattened by the growing non-canonical kinetic term, which also makes the field range of the physical field unbounded.
As a consequence, any combination of mass and decay constant can generate enough ALP energy density to account for all the dark matter that we observe in the universe. 

An important cosmological constraint arises from isocurvature fluctuations imprinted by inflation. 
To apply these constraints to our scenario we give a simple derivation of the size of isocurvature fluctuations in general models with arbitrary potential and even a temperature dependence of the potential.
As a useful crosscheck we have updated the isocurvature constraints~\autocite{visinelli_dark_2009, kobayashi_isocurvature_2013} using the newest Planck data~\autocite{ade_planck_2016} and the most recent results for the QCD topological susceptibility~\autocite{borsanyi_axion_2016}.
The result can be found in \figref{fig:QCD_isocurvature_cosine}.
In our non-canonical setup the isocurvature constraints are even slightly weaker as can be seen in \figref{fig:QCD_isocurvature_noncan}.

An interesting non-trivial situation arises if the ALP is coupled to the strong interactions, i.e. via a term $\sim\phi G\tilde{G}$.
This is of particular interest since a number of experiments are currently searching for ALP dark matter with this coupling~\autocite{abel_search_2017, budker_cosmic_2014,hexenia_2017}.
The coupling to gluons leads to two non-trivial features: the generation of a temperature-dependent, irreducible contribution to the ALP mass and an effective ALP field value dependent nucleons mass.
The former naively makes large parts of the low mass region explored by current experiments inaccessible~\autocite{blum_constraining_2014}.
This can be avoided by invoking a precise cancellation with an additional term in the ALP potential (with or without non-canonical terms).
The latter leads to strong constraints from Big Bang Nucleosynthesis.
These are significantly weakened in our scenario with a non-canonical kinetic term.
This opens up significant parameter space that can be explored in near future experiments such as Casper~\autocite{budker_cosmic_2014} and HeXeniA~\autocite{hexenia_2017}, as well as EDM storage rings~\autocite{chang_axion_2017}.

\section*{Acknowledgements}
This project has received funding from the European Union's Horizon 2020 research and innovation programme under the Marie Sklodowska-Curie grant agreements No $690575$ (RISE InvisiblesPlus) and No $674896$ (ITN ELUSIVES).

%%%%%%%%%%%%%%%%%%%%%%%%%%%%%%%%%%%%%%%%%%%%%%%%
\appendix
\section{Effect of higher-order poles in the kinetic function}\label{sec:Appendix1}

In this appendix we briefly study how our results change if we allow our non-canonical kinetic term for the ALP field to have a pole of arbitrary (even) order.
We work with the Lagrangian~\eqref{eq:basic_lagrangian}, this time with the kinetic function given by
\begin{equation}
K(\phi) = \frac{1}{\cos^{p} \left( \frac{N\phi}{f_a} \right)}\ ,\quad p\in\mathbb{N}.
\end{equation}
Note that with this definition the order of the pole is $2p$.
In the main body we have focused in the $p=1$ case.
As opposed to the $p=1$ case, for a general value of $p$ it is not possible to find an exact analytic expression for the transformation to the canonically normalised field $\psi(\phi)$.
However, we can find an approximate expression, valid close to the pole at $\phi/f_a = \pi/(2N)$, by expanding $K(\phi)$ in a Laurent series and keeping only the leading divergent term.
With that, we can then proceed as in~\secref{sec:analytic_estimate} and obtain an estimation for the enhancement in the relic density.
The result is
\begin{equation}\label{eq:p_cases}
\frac{\rho^{\mathrm{anh}}}{\rho^{\mathrm{harm}}} \propto 
\left\{
\begin{aligned}
&\exp^{\frac{3}{4}N\psi_0},\quad & p&=1 \\
&\left( N\psi_0 \right)^{\frac{3}{2}\left(1+\frac{1/2}{k-1}\right)},\quad & p&\neq1
\end{aligned}
\right.
.
\end{equation}
Due to the now unbounded field range, a significant enhancement is possible.
Nevertheless, it may seem that the enhancement effect is much weaker in the $p>1$ cases, which could seem counter-intuitive.
However, we must note that the field redefinition $\psi(\phi)$ is different for different values of $p$, which means that it is not so obvious to compare the distinct cases just by looking at~\eqref{eq:p_cases}.
In order to be able to compare, we can recast~\eqref{eq:p_cases} in terms of the non-canonically normalised field $\phi$, which avoids the problem of the $p$-dependent field redefinition.
Doing so, we can write
\begin{equation}\label{eq:p_cases_theta}
\frac{\rho^{\mathrm{anh}}}{\rho^{\mathrm{harm}}} \propto \left( \frac{1}{\sqrt{\frac{\pi}{2N}-\theta_0}} \right)^{2p-1},
\end{equation}
which is valid for all $p\in \mathbb{N}$.
Looking at~\eqref{eq:p_cases_theta} we can confirm that there exists an enhancement in the ALP relic density for all values of $p$.
What's more, looking at it from the point of view of the non-canonically normalised field, the effect is stronger the higher the order of the pole, as one would naively expect.

\section{Temperature dependent anharmonicity function}\label{sec:Appendix2}

In this section we detail how to implement the effects of the temperature dependence of the QCD axion mass into the anharmonicity function~\autocite{visinelli_dark_2009,diez-tejedor_cosmological_2017}.
The approach that we follow allows us to analytically upgrade any anharmonicity function that does not include temperature effects into a full temperature dependent anharmonicity function.
The derivation that we present is valid for any scalar field whose potential can be factorised as
\begin{equation}
V(\phi) = m^2(T)\cdot V_0(\phi),
\end{equation}
where $V_0(\phi)$ is a temperature-independent potential that has a minimum around which the field can oscillate (possibly anharmonically), and the temperature dependence acts only as a scaling.
This is the case for general axion models, including those we study in this paper.
If there is no temperature dependence at all, then $m(T)\equiv m$ and working as in~\secref{sec:numerics} we can express the energy density of the field with an anharmonic potential as
\begin{equation}
\rho^{\mathrm{anh}} = f(\theta_0) \rho^{\mathrm{harm}},
\end{equation}
where $\rho^{\mathrm{harm}}$ is the solution to the harmonic case described in~\secref{sec:misalignment} and $f(\theta_0)$ is the anharmonicity function that depends on the initial value of the dimensionless field $\theta=\phi/f_a$.

At the effective level, we can think that the only effect of the anharmonicities is to change the time (or temperature) at which the oscillations start.
This is of course not what actually happens, but with this approach we will be able to make a good estimate of the energy density of the field at late times.
In this picture, we have that
\begin{equation}\label{eq:anharmonicity_start_temperature}
f(\theta_0) = \frac{\rho^{\mathrm{anh}}}{\rho^{\mathrm{harm}}} = \left( \frac{T_\mathrm{S}^{\mathrm{harm}}}{T_\mathrm{S}^{\mathrm{anh}}} \right) ^3 \frac{g_{\star S}(T_\mathrm{S}^\mathrm{harm})}{g_{\star S}(T_\mathrm{S}^{\mathrm{anh}})},
\end{equation}
which follows from the dependence of the WKB solution \eqref{eq:energy_density_semianalytic} on the temperature at which oscillations start\footnote{In the following equations, we will neglect all instances of quotients of effective degrees of freedom, as they only introduce a small correction and make the derivation much more cumbersome.}.
With this we can extend the equation for the condition of the start of the oscillations in the harmonic case, $3H(T_\mathrm{S}^{\mathrm{harm}})=m$, to the anharmonic case, using an effective mass that encodes the effects of the anharmonicity.
It reads
\begin{equation}\label{eq:anh_start}
3H(T_\mathrm{S}^{\mathrm{anh}})=m\left(f(\theta_0)\right)^{-2/3}.
\end{equation}
This expression will be useful later on.

We now assume that there is a temperature dependent mass that evolves as \eqref{eq:QCD_axion_mass}, as is the case for axion models.
As was argued in the main body of the paper, if $T_\mathrm{S}^{\mathrm{anh}} < T_\mathrm{crit}$, the oscillations start after the QCD phase transition, when the mass has already attained its low-temperature value, and the temperature dependence has no effect on the later evolution of the field.
In the harmonic case, this happens if the zero-temperature mass is smaller than $m^*$ given by $3H(T_\mathrm{crit})=m^*$, that is,
\begin{equation}
m^* = 3\cdot 1.66\sqrt{g_\star(T_\mathrm{crit})} \frac{T_\mathrm{crit}^2}{m_\mathrm{Pl}} \simeq 6.6\cdot 10^{-11}\ \mathrm{eV}.
\end{equation}
In terms of decay constants, this translates into a maximum value $f_a^*  \simeq 8.7\cdot 10^{16}\,\mathrm{GeV}$ above which the temperature dependence does not play a role.
In the anharmonic case, it can happen that the anharmonicities delay the start of the oscillations beyond $T_\mathrm{crit}$, even if $f_a<f_a^*$, if the initial misalignment angle is large enough.
We can use equation \eqref{eq:anharmonicity_start_temperature} to find an expression for this critical value $\theta_0^*$.
Writing it in terms of $f_a^*$, it reads
\begin{equation}
\left( \frac{f_a^*}{f_a} \right)^{3/2} = f(\theta_0^*),
\end{equation}
from where $\theta_0^*$ can be obtained once an explicit anharmonicity function is chosen.

Finally, if both $f_a<f_a^*$ and $\theta_0<\theta_0^*$, then $T_\mathrm{S}^{\mathrm{anh}} > T_\mathrm{crit}$ and the temperature dependent evolution of the mass will have an impact on the oscillating behaviour of the field.
First of all, we compute how the onset of the oscillations is modified.
For this purpose, we can just substitute the constant mass $m$ for the temperature dependent one $m(T)$ in equation \eqref{eq:anh_start}.
Using the expression for $m(T)$ given in \eqref{eq:QCD_axion_mass}, we have the condition
\begin{equation}
3H(T_\mathrm{S})=m\left( \frac{T_\mathrm{S}}{T_\mathrm{crit}} \right)^\alpha \left(f(\theta_0)\right)^{-2/3}.
\end{equation}
To simplify the notation, we have denoted $T_\mathrm{S}$ the temperature at which the oscillations start if we take into account both the anharmonicities of $V_0(\phi)$ and the temperature dependence of $m(T)$.
We can recast this equation in terms of decay constants, finding
\begin{equation}
\frac{T_\mathrm{S}}{T_\mathrm{crit}} = \left( \frac{f_a^*}{f_a}\cdot\left(f(\theta_0)\right)^{-2/3} \right)^{\frac{1}{2-\alpha}}.
\end{equation}

To continue, we use the expression for the energy density of an oscillating scalar field with a slowly varying mass, which can be found for instance in~\autocite{arias_wispy_2012} and reads\footnote{This expression is a generalisation of the WKB approximation presented before.}
\begin{equation}
\rho(T) = \frac{1}{2} m(T) m(T_\mathrm{S}) f_a^2\theta_0^2\, \frac{g_{\star S}(T)}{g_{\star S}(T_\mathrm{S})} \left( \frac{T}{T_\mathrm{S}} \right)^3.
\end{equation}
At low temperatures below the QCD critical temperature, the quotient between this expression and the corresponding one for the harmonic case is
\begin{equation}
\frac{\rho(T)}{\rho^\mathrm{harm}(T)} = \frac{m(T_\mathrm{S})}{m} \frac{g_{\star S}(T^\mathrm{harm}_\mathrm{S})}{g_{\star S}(T_\mathrm{S})} \left( \frac{T^\mathrm{harm}_\mathrm{S}}{T_\mathrm{S}} \right)^3.
\end{equation}
But this is precisely what we need to define the temperature dependent anharmonicity function $F(\theta_0,f_a)$.
Again, neglecting the quotient of effective degrees of freedom, we find
\begin{equation}
\begin{aligned}
F(\theta_0,f_a) &\equiv \frac{\rho(T)}{\rho^\mathrm{harm}(T)} \\
& \simeq \frac{m(T_\mathrm{S})}{m} \left( \frac{T^\mathrm{harm}_\mathrm{S}}{T_\mathrm{crit}} \right)^3 \left( \frac{T_\mathrm{crit}}{T_\mathrm{S}} \right)^3 \\
& = \left( \frac{T_\mathrm{S}}{T_\mathrm{crit}} \right)^{\alpha-3} \left( \frac{f_a}{f_a^*} \right)^{-3/2} \\
& = \left( \frac{f_a^*}{f_a} \right)^{\frac{\alpha}{2(2-\alpha)}} \left( f (\theta_0) \right)^{\frac{2(3-\alpha)}{3(2-\alpha)}}.
\end{aligned}
\end{equation}
It can be checked that this result agrees with the ones given in~\autocite{visinelli_dark_2009,diez-tejedor_cosmological_2017}, but can be applied in more general contexts.

\printbibliography

\end{document}